\global\def\draftcontrol{0}

   \def\versionno{ General BCFW -- draft   }

\catcode`\@=11

\expandafter\ifx\csname draftcontrol\endcsname\relax\global\def\draftcontrol{0}
\fi

{\count255=\time\divide\count255 by 60
\xdef\hourmin{\number\count255}
\multiply\count255 by-60\advance\count255 by\time
\xdef\hourmin{\hourmin:\ifnum\count255<10 0\fi\the\count255}}
\def\draftdate{\number\month/\number\day/\number\year\ \ \ \hourmin }

\newcommand\makepapertitle{\par
  \begingroup
    \renewcommand\thefootnote{\@fnsymbol\c@footnote}%
    \def\@makefnmark{\rlap{\@textsuperscript{\normalfont\@thefnmark}}}%
    \long\def\@makefntext##1{\parindent 1em\noindent
            \hb@xt@1.8em{%
                \hss\@textsuperscript{\normalfont\@thefnmark}}##1}%
     \newpage
     \global\@topnum\z@   
     \@makepapertitle
     \thispagestyle{empty}\@thanks
  \endgroup
  \setcounter{footnote}{0}%
  \global\let\thanks\relax
  \global\let\makepapertitle\relax
  \global\let\@makepapertitle\relax
  \global\let\@thanks\@empty
  \global\let\@author\@empty
  \global\let\@date\@empty
  \global\let\@title\@empty
  \global\let\title\relax
  \global\let\author\relax
  \global\let\date\relax
  \global\let\and\relax
  \def\version{\let\version\@version\@gobble}
}
\def\@makepapertitle{%
  \newpage
   \ifnum\draftcontrol=1 {}
   \version\versionno
   \vskip 3em%
   \else
   \hfill\hbox to 3cm {\parbox{4cm}{\@pubnum}\hss}%
   \vskip 3em%
   \fi
   \begin{center}%
   \let \footnote \thanks
     {\LARGE {\@title}}%
     \vskip 1.5em%
     {\normalsize
       \lineskip .5em%
       \begin{tabular}[t]{c}%
         \@author
       \end{tabular}\par}%
     \vskip 1.5em%
     {\@bstract}%
     \end{center}%
     \vskip 1.5em
     \@date%
   \par
}

\gdef\@pubnum{}
\def\pubnum#1{%
  \gdef\@pubnum{#1}}

\gdef\@bstract{}
\def\Abstract#1{%
  \gdef\@bstract{%
   \parbox{\textwidth-0pc}{%
   \centerline{\bf Abstract}\penalty1000%
\kern.2cm%
\noindent
\renewcommand\baselinestretch{1.0}%
{#1}}}
}

\def\ps@paper{\let\@mkboth\@gobbletwo%
     \ifnum\draftcontrol=1
    \def\@oddfoot{\hbox to \textwidth{\tiny \versionno \hfil\tiny\draftdate}%
    \hskip -\textwidth \hbox to \textwidth{\hfil\rm\thepage\hfil}}%
     \else\def\@oddfoot{\hbox to \textwidth{\hfil\rm\thepage\hfil}}
     \fi
     \let\@evenfoot\@oddfoot
}

\def\body{\clearpage
          \pagestyle{paper}
    }

\def\@version#1{\ifnum\draftcontrol=1
\typeout{}\typeout{#1}\typeout{}
\vskip3mm\centerline{\hbox{\fbox{\normalsize{\tt DRAFT -- #1 -- }
                   {\draftdate}}}}\vskip3mm
\fi}
\let\version\@version
\long\def\eqlabel#1{\ifnum\draftcontrol=1
                    \tag@false  
                    \tag*{(\theequation) \hbox to -0.2cm{\hspace{0cm}\small{#1}\hss}}
                    \refstepcounter{equation}
                    \edef\@currentlabel{\theequation}
                    \ltx@label{#1}          
                    \else
                    \label{#1}
                    \fi
                    }
\let\st@bibitem\@bibitem
\let\st@lbibitem\@lbibitem
\ifnum\draftcontrol=1
  \def\@bibitem#1{%
    \st@bibitem{#1}\a@@label{#1}\ignorespaces}
  \def\@lbibitem[#1]#2{%
    \st@lbibitem[#1]{#2}\a@@label{#2}\ignorespaces}
  \def\a@@label#1{%
    \gdef\a@lab{\smash{\normalfont\small#1}}
    \ifvmode
      \if@inlabel
        \global\setbox\@labels\hbox{%
          \llap{\a@lab\let\a@lab\relax
                \kern\@totalleftmargin\kern\marginparsep}%
          \box\@labels}%
      \fi
    \fi}
\fi

\documentclass[11pt,letterpaper]{article}

\usepackage{amsmath,amssymb,array,calc,epsfig}
\usepackage{cite}
\usepackage[
      colorlinks=true,
      linkcolor=red,  
      urlcolor=blue,    
      filecolor=red,     
      linkcolor=blue,
      citecolor = red,
      pdfstartview=FitV,
      pdftitle={},
        pdfauthor={Mukund Rangamani},
        pdfsubject={},
        pdfkeywords={},
        pdfpagemode=None,
        bookmarksopen=true    
      ]{hyperref}
      
\usepackage{graphicx}
\usepackage{subfigure}
\usepackage{epstopdf}
\usepackage{ccaption}

\ifnum\draftcontrol=1
\fi

\renewcommand\baselinestretch{1.25}
\renewcommand\section{\@startsection {section}{1}{\z@}%
                                   {-3.5ex \@plus -1ex \@minus -.2ex}%
                                   {2.3ex \@plus.2ex}%
                                   {\normalfont\large\bfseries}}
\renewcommand\subsection{\@startsection{subsection}{2}{\z@}%
                                   {-3.25ex\@plus -1ex \@minus -.2ex}%
                                   {1.5ex \@plus .2ex}%
                                   {\normalfont\normalsize\bfseries}}
\renewcommand\subsubsection{\@startsection{subsubsection}{3}{\z@}%
                                   {-3.25ex\@plus -1ex \@minus -.2ex}%
                                   {1.5ex \@plus .2ex}%
                                   {\normalfont\normalsize\it}}
\renewcommand\paragraph{\@startsection{paragraph}{4}{\z@}%
                                   {-3.25ex\@plus -1ex \@minus -.2ex}%
                                   {1.5ex \@plus .2ex}%
                                   {\normalfont\normalsize\bf}}

\numberwithin{equation}{section}

\def\revise#1       {\raisebox{-0em}{\rule{3pt}{1em}}%
                     \marginpar{\raisebox{.5em}{\vrule width3pt\
                     \vrule width0pt height 0pt depth0.5em
                     \hbox to 0cm{\hspace{0cm}{%
                     \parbox[t]{4em}{\raggedright\footnotesize{#1}}}\hss}}}}

\def\sqr#1#2{{\vcenter{\vbox{\hrule height.#2pt
 \hbox{\vrule width.#2pt height#1pt \kern#1pt
 \vrule width.#2pt}\hrule height.#2pt}}}}

\def\aa1{\phi}
\def\cc1{\psi}

\catcode`\@=12

\begin{document}


\title{\bf On the Tree-Level Structure of Scattering Amplitudes of Massless Particles}

\pubnum{%
arXiv:xxxx.xxxx}
\date{xxxxx}

\author{
\scshape Paolo Benincasa${}^{\dagger}$, Eduardo Conde$^{\ddagger}$\\[0.4cm]
\ttfamily Departamento de F{\'i}sica de Part{\'i}culas, Universidade de Santiago de Compostela\\
\ttfamily and\\
\ttfamily Instituto Galego de F{\'i}sica de Altas Enerx{\'i}as (IGFAE)\\
\ttfamily E-15782 Santiago de Compostela, Spain\\[0.2cm]
\small \ttfamily ${}^{\dagger}$paolo.benincasa@usc.es, ${}^{\ddagger}$eduardo@fpxp1.usc.es
}

\Abstract{We provide a new set of on-shell recursion relations for tree-level scattering amplitudes, which are 
valid for any non-trivial theory of massless particles. In particular, we reconstruct the scattering 
amplitudes from (a subset of) their poles and zeroes. The latter determine the boundary term arising in the 
BCFW-representation when the amplitudes do not vanish as some momenta are taken to infinity along some complex 
direction. Specifically, such a boundary term can be expressed as a sum of products of two on-shell amplitudes
with fewer external states and a factor dependent on the location of the relevant zeroes and poles. This
allows us to recast the amplitudes to have the standard BCFW-structure, weighted by a simple factor
dependent on a subset of zeroes and poles of the amplitudes. We further comment on 
the physical interpretation of the zeroes as a particular kinematic limit in the complexified momentum 
space. The main implication of the existence of such recursion relations is that the tree-level approximation 
of any consistent theory of massless particles can be fully determined just by the knowledge of the corresponding
three-particle amplitudes.}

\makepapertitle

\body

\version\versionno

\section{Introduction}\label{sec:Intro}

In recent years a great deal of progress has been made in understanding the 
perturbation theory for massless particles. In particular, a class of theories, named {\it fully 
constructible} \cite{Benincasa:2007xk}, was found to be characterised by tree-level scattering amplitudes 
satisfying on-shell recursion relations 
\cite{Britto:2004aa, Britto:2005aa, Bedford:2005yy, Benincasa:2007qj, ArkaniHamed:2008yf, 
Cheung:2008dn, ArkaniHamed:2008gz}, 
{\it i.e.} amplitudes with an arbitrary number of external states are expressed in terms of on-shell 
amplitudes with fewer external states. Iterating such recursion relations, an $n$-particle amplitude
turns out to be determined just in terms of three-particle amplitudes, which are non-vanishing in the complexified
momentum space \cite{Witten:2003aa}, irrespectively of the vertex structure in the Feynman representation.
This result is incredibly striking, especially if one keeps in mind that in the Feynman representation
an $n$-particle amplitude may receive contributions from all the $k$-point vertices with $k\,\le\,n$. 

There is a very general method (BCFW-construction) \cite{Britto:2005aa} which can point out the presence of such a 
structure. The idea is to introduce a one-parameter deformation of the complexified momentum space, which 
generates a one-parameter family of amplitudes, and try to reconstruct the physical amplitude from the 
singularity structure of this family as a function of the deformation parameter. Restricting the
analysis to the pole structure is equivalent to considering the tree-level approximation. Thus, reconstructing 
scattering amplitudes from their pole structure means relating them to the residues of such poles. This is 
indeed possible if the deformed amplitudes vanish as momenta are taken to infinity along some complex
direction, as it was shown in \cite{Britto:2005aa} for an arbitrary number of external gluons, and in 
\cite{Benincasa:2007qj} for an arbitrary number of external gravitons. The residues are just products of
two on-shell scattering amplitudes with fewer external states. An on-shell amplitude may therefore be related 
to on-shell amplitudes with fewer external particles, providing a recursion relation. 

In the case that the amplitude does not vanish as momenta are taken to infinity along some complex direction, one would need 
to consider a further contribution besides the ones coming from the poles at finite locations. Such a boundary term has not
been understood, beyond some small steps made in \cite{Feng:2009ei, Feng:2010ku}, or the possibility of setting it to zero 
with some particular gauge choice \cite{Bianchi:2008pu}. It would be desirable to have a deeper understanding of it, and it
will be the main subject of this paper. 

\begin{figure}[htbp]
 \centering 
  \scalebox{.30}{\includegraphics{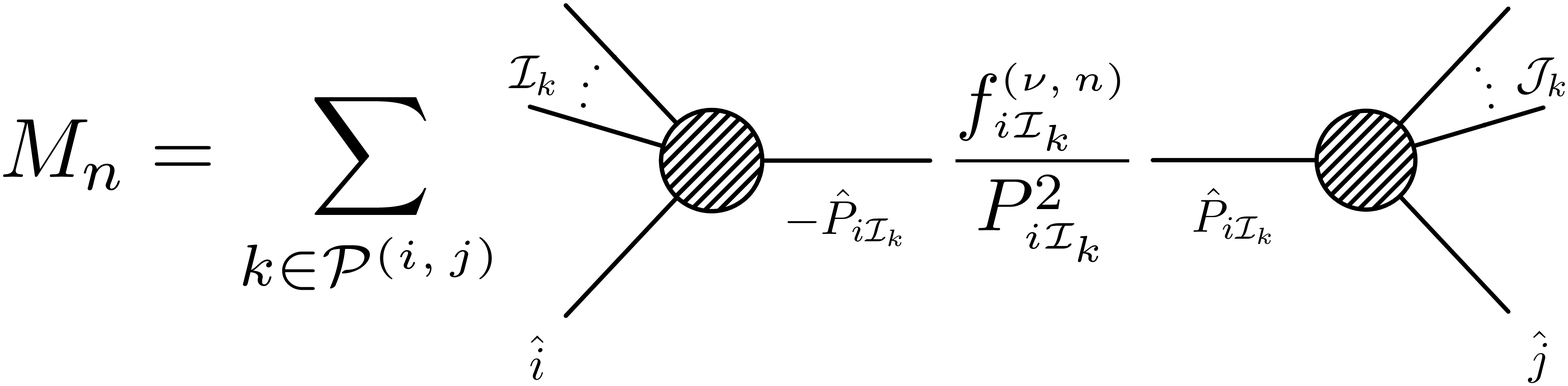}}
  \caption{Generalised on-shell recursion relation. This new recursion relation shows the same structure of the usual
           BCFW one, with a further factor which depends on a sub-set of the zeroes of the amplitude.}\label{fig:wBCFW}
\end{figure}

Let us anticipate here the most striking results. First, we claim that all theories are tree-level constructible.
Specifically, we prove the existence of generalised on-shell recursion relations with the same structure as the standard
BCFW ones (figure \ref{fig:wBCFW}) by proposing to consider a subset of zeroes of the amplitudes

\begin{equation}\eqlabel{Mfin2}
 M_n\:=\:\sum_{k\in\mathcal{P}^{\mbox{\tiny $(i,j)$}}}
  M_{\mbox{\tiny L}}^{\mbox{\tiny $(i,j)$}}(\hat{i},\mathcal{I}_k,-\hat{P}_{\mbox{\tiny $i\mathcal{I}_k$}})
  \frac{f_{\mbox{\tiny $i\mathcal{I}_k$}}^{\mbox{\tiny $(\nu,n)$}}}{P_{\mbox{\tiny $i\mathcal{I}_k$}}^2}
  M_{\mbox{\tiny R}}^{\mbox{\tiny $(i,j)$}}(\hat{P}_{\mbox{\tiny $i\mathcal{I}_k$}}, \mathcal{J}_{k}, \hat{j}),
\end{equation}
with the factors $f_{\mbox{\tiny $i\mathcal{I}_k$}}^{\mbox{\tiny $(\nu,n)$}}$ being
\begin{equation}\eqlabel{fweight}
 f_{\mbox{\tiny $i\mathcal{I}_k$}}^{\mbox{\tiny $(\nu,n)$}}\:=\:
 \left\{
  \begin{array}{l}
   1,\hspace{4cm} \nu\,<\,0,\\
   \phantom{\ldots}\\
   \prod_{l=1}^{\nu+1}\left(1-\frac{P_{\mbox{\tiny $i\mathcal{I}_k$}}^2}{P_{\mbox{\tiny $i\mathcal{I}_k$}}^2
    \left(z_0^{\mbox{\tiny $(l)$}}\right)}\right),\quad \nu\,\ge\,0,
  \end{array}
 \right.
\end{equation}
where the Lorentz invariants computed at the location of the zeroes $\left\{z_0^{\mbox{\tiny $(l)$}}\right\}$ are
constrained by unitarity, and $\nu$ characterises the large-$z$ behaviour of the amplitude 
($M_{n}(z)\,\overset{\mbox{\tiny $z\rightarrow\infty$}}{\sim}\,z^{\nu}$). The form of eq \eqref{Mfin2} 
extends the notion of constructibility to all theories of massless particles.

Furthermore, the analysis of the collinear and multi-particle limits of the representation \eqref{Mfin2} will allow us to 
prove that the behaviour at infinity of the amplitude does not depend on the number of external particles, but it rather 
depends on the number of derivatives $\delta$ of the three-particle interaction as well as on the helicities of the 
particles whose momenta have been deformed.

The paper is organised as follows. In section \ref{sec:BCFW} we review the BCFW construction both in general and its
realisation using the four-dimensional spinor-helicity formalism. In section \ref{sec:Bound} we discuss the limit of
the amplitudes as the momenta are taken to infinity along some complex direction, the degree at which it might diverge,
the eventual contribution from the boundary term and their relation with the zeroes of the amplitudes. Here we derive as
well the generalised recursion relation. In section \ref{sec:ZeroLimits} we provide the conditions on the channels
computed at the location of the zeroes by requiring that our recursion relation satisfies the correct collinear and
multi-particle limits. section \ref{sec:complexUV} is dedicated to further comments on the large-$z$ behaviour. 
In section \ref{sec:Ex}, we work out some examples. Finally, section \ref{sec:Concl} contains the conclusion and further 
discussions.

\section{BCFW Construction}\label{sec:BCFW}

In the complexified momentum space it is possible to introduce a one-complex parameter deformation such that the deformed 
momenta satisfy both the on-shell condition and the momentum conservation \cite{Britto:2005aa}, defining a one-parameter 
family of amplitudes. This deformation is certainly not unique. The simplest example may be defined by deforming the momenta
of just two particles, leaving the others unchanged
\begin{equation}\eqlabel{BCFWdef}
 p^{\mbox{\tiny $(i)$}}(z)\:=\:p^{\mbox{\tiny $(i)$}}-zq,\qquad
 p^{\mbox{\tiny $(j)$}}(z)\:=\:p^{\mbox{\tiny $(j)$}}+zq,\qquad
 p^{\mbox{\tiny $(k)$}}(z)\:=\:p^{\mbox{\tiny $(k)$}},
\end{equation}
where $k$ labels all the particles except for $i$ and $j$. The deformation \eqref{BCFWdef} straightforwardly
satisfies momentum conservation. The requirement that the deformed momenta are on-shell fixes $q$ to be
necessarily complex and such that
\begin{equation}\eqlabel{q}
 q^2\:=\:0,\qquad p^{\mbox{\tiny $(i)$}}\,\cdot\,q\:=\:0\:=\:p^{\mbox{\tiny $(j)$}}\,\cdot\,q.
\end{equation}
A deformation of this type defines a one-parameter family of amplitudes $M_n(z)$, and it is now possible to
analyse the singularity structure of the amplitude as a function of $z$. The poles in $z$ are located in
correspondence with the $z$-dependent propagators and they turn out to be all simple poles
\begin{equation}\eqlabel{poles}
 \frac{1}{\left[P_{\mbox{\tiny $k$}}(z)\right]^2}\:=\:
  \frac{1}{P_{\mbox{\tiny $k$}}^2-2z\left(P_{\mbox{\tiny $k$}}\cdot q\right)}\:\Rightarrow\:
  z_{\mbox{\tiny $k$}}\:=\:\frac{P_{\mbox{\tiny $k$}}^2}{2P_{\mbox{\tiny $k$}}\cdot q},
\end{equation}
where $P_{\mbox{\tiny $k$}}\,=\,\sum_{l\in\mathcal{S}_k} p^{\mbox{\tiny $(l)$}}$, $\mathcal{S}_k$ is the set of the 
momenta of the external particles in the $k$-channel, $i\,\in\,\mathcal{S}_k$ and 
$z_{\mbox{\tiny $k$}}$ is the location of the pole. As $z\,\rightarrow\,z_{\mbox{\tiny $k$}}$ the momentum
$P_{\mbox{\tiny $k$}}$ in \eqref{poles} goes on-shell and this channel factorises into the product of 
two on-shell amplitudes
\begin{equation}\eqlabel{res}
 M_{n}^{\mbox{\tiny $(i,j)$}}(z)\:\overset{\mbox{\tiny $z\rightarrow z_{\mbox{\tiny $k$}}$}}{\sim}\:
  \frac{M_{\mbox{\tiny L}}^{\mbox{\tiny $(i,j)$}}(z_k)M_{\mbox{\tiny R}}^{\mbox{\tiny $(i,j)$}}(z_k)}{
  P^2_{\mbox{\tiny $k$}}(z)},
\end{equation}
where the notation $M_{n}^{\mbox{\tiny $(i,j)$}}(z)$ just indicates that the one-parameter family of 
amplitudes has been obtained by deforming the momenta of the particles labelled by $i$ and $j$.
The physical amplitude, which can be obtained from $M_{n}(z)$ by setting $z\,=\,0$, is related to the 
residues of the poles in \eqref{res}. Let $\mathcal{R}$ be the
Riemann sphere obtained as union of the complex plane with the point at infinity 
$\mathcal{R}\,=\,\mathbb{C}\cup\left\{\infty\right\}$, then
\begin{equation}\eqlabel{AmplInt}
  0\:=\:\frac{1}{2\pi i}\oint_{\mbox{\tiny $\mathcal{R}$}}\frac{dz}{z}M_{n}^{\mbox{\tiny $(i,j)$}}(z)\:=
     \:M_{n}(0)-\sum_{k\in\mathcal{P}^{\mbox{\tiny $(i,j)$}}}
    \frac{M_{\mbox{\tiny L}}^{\mbox{\tiny $(i,j)$}}(z_k)M_{\mbox{\tiny R}}^{\mbox{\tiny $(i,j)$}}(z_k)}{P^2_k}
     -\mathcal{C}_{n}^{\mbox{\tiny $(i,j)$}}.
\end{equation}
If $M_{n}^{\mbox{\tiny $(i,j)$}}(z)\,\overset{\mbox{\tiny $z\rightarrow\infty$}}{\rightarrow}0$, the boundary term 
$\mathcal{C}_{n}^{\mbox{\tiny $(i,j)$}}$ is zero and the solution of eq \eqref{AmplInt} for $M_{n}(0)$
provides with a recursion relation. Such a condition is satisfied by Yang-Mills \cite{Britto:2005aa},
gravity \cite{Benincasa:2007qj}, $\mathcal{N}=4$ Super Yang-Mills and $\mathcal{N}=8$ Supergravity
\cite{ArkaniHamed:2008gz}, and for amplitudes with external gravitons/gluons, as highest spin particles, under
a deformation of the momenta of the gravitons/gluons \cite{Cheung:2008dn}.

\subsection{The BCFW Construction and the Spinor-Helicity Formalism}

In this section, we briefly introduce the spinor-helicity representation of the scattering amplitudes, 
that allows to express them as a function of a set of pairs of spinors 
$\left\{\left(\lambda^{\mbox{\tiny $(i)$}},\,\tilde{\lambda}^{\mbox{\tiny $(i)$}}\right)\right\}$ and
of the helicities $\left\{h_i\right\}$ of the particles. In four dimensions, this equivalence holds because
of the isomorphism $SO(3,1)\,\cong\,SL(2,\mathbb{C})$, which is implemented by the Pauli matrices
$\sigma^{\mu}_{a\dot{a}}\,=\,\left(\mathbb{I}_{a\dot{a}},\,\overrightarrow{\sigma}_{a\dot{a}}\right)$:
\begin{equation}\eqlabel{iso}
 p_{\mu}\quad\longrightarrow\quad p_{a\dot{a}}\:=\:\sigma^{\mu}_{a\dot{a}}p_{\mu}
  \:=\:\lambda_{a}\tilde{\lambda}_{\dot{a}},
\end{equation}
with $\lambda_{a}$ and $\tilde{\lambda}_{\dot{a}}$ transforming respectively in the $(1/2,\,0)$ and 
$(0,\,1/2)$ representation of $SL(2,\mathbb{C})$. In the real momentum space, these two spinors are related by
complex conjugation: $\tilde{\lambda}_{\dot{a}}\,=\,\left(\lambda_a\right)^{\star}$. 

It is possible to define two inner product for spinors, one for each representation of $SL(2,\mathbb{C})$ under which they
can transform:
\begin{equation}\eqlabel{innprod}
 \langle\lambda,\,\lambda'\rangle\:\equiv\:\epsilon^{ab}\lambda_{a}\lambda'_{b},\qquad
 [\tilde{\lambda},\,\tilde{\lambda}']\:\equiv\:\epsilon^{\dot{a}\dot{b}}
  \tilde{\lambda}_{\dot{a}}\tilde{\lambda}'_{\dot{b}},
\end{equation}
with $\epsilon_{12}\,=\,1\,=\,\epsilon_{\dot{1}\dot{2}}$, $\epsilon^{12}\,=\,-1\,=\,\epsilon^{\dot{1}\dot{2}}$,
and $\epsilon^{ac}\epsilon_{cb}\,=\,\delta^{a}_{\phantom{a}b}$. Notice that the inner products \eqref{innprod}
are Lorentz invariant. 

Assuming the existence of one-particle states and that the Poincar{\'e} group acts on scattering amplitudes
as it acts on individual states, the helicity operator acts on the amplitudes as follows:
\begin{equation}\eqlabel{helop}
 \left(\lambda^{\mbox{\tiny $\left(i\right)$}}_{a}
  \frac{\partial}{\partial\lambda^{\mbox{\tiny $\left(i\right)$}}_{a}}-
 \tilde{\lambda}^{\mbox{\tiny $\left(i\right)$}}_{\dot{a}}
  \frac{\partial}{\partial\tilde{\lambda}^{\mbox{\tiny $\left(i\right)$}}_{\dot{a}}}
 \right)M_{n}\left(1^{\mbox{\tiny $h_{i}$}},\ldots,n^{\mbox{\tiny $h_{i}$}}\right)\:=\:-2h_{i}\,
  M_{n}\left(1^{\mbox{\tiny $h_{i}$}},\ldots,n^{\mbox{\tiny $h_{i}$}}\right).
\end{equation}

In the complexified Minkowski space, the isometry group is 
$SO(3,1,\mathbb{C})\,\cong\,SL(2,\mathbb{C})\,\times\,SL(2,\mathbb{C})$ and each of the two spinors 
$\lambda_{\mbox{\tiny $a$}}$ and $\tilde{\lambda}_{\mbox{\tiny $\dot{a}$}}$ can be taken to belong to a different
copy of $SL(2,\mathbb{C})$, so that they are not related by complex conjugation and, therefore, are independent of
each other. Working in the complexified momentum space, the BCFW-deformation \eqref{BCFWdef} in terms of the spinors 
\eqref{iso} turns out to be
\begin{equation}\eqlabel{BCFWdef2}
 \tilde{\lambda}^{\mbox{\tiny $(i)$}}(z)\:=\:\tilde{\lambda}^{\mbox{\tiny $(i)$}}-z\tilde{\lambda}^{\mbox{\tiny $(j)$}},
 \quad
 \lambda^{\mbox{\tiny $(j)$}}(z)\:=\:\lambda^{\mbox{\tiny $(j)$}}+z{\lambda}^{\mbox{\tiny $(i)$}}.
\end{equation}
The expression \eqref{BCFWdef2} actually defines a set of four deformations, which are determined by the sign of the
helicities $\left(h_i,\,h_j\right)$ of the deformed particles: 
\begin{equation}\eqlabel{BCFWhel}
 \left(h_i,\,h_j\right)\:=\:\left\{\left(-,\,+\right), \left(-,\,-\right), \left(+,\,+\right), \left(+,\,-\right)\right\}.
\end{equation}
Depending on the helicity configuration \eqref{BCFWhel} chosen for the two-particle deformation \eqref{BCFWdef2},
the induced one-parameter family of amplitudes is a different function of $z$ and, therefore, the boundary term
$\mathcal{C}_{n}^{\mbox{\tiny $(i,j)$}}$ differs as well.

As mentioned earlier, the helicity-spinor formalism as discussed here is strictly four-dimensional. Generalisations
have been proposed in \cite{Cheung:2009dc, Boels:2009bv, CaronHuot:2010rj} for higher dimensions and in \cite{Gang:2010gy}
for three-dimensions. Our main results heavily rely on the four-dimensional helicity-spinor formalism. However, as showed
in \cite{ArkaniHamed:2008yf}, the BCFW-structure of the tree level amplitude is a general feature of a theory and is
not related to the dimensionality of the space-time. Therefore, in order to generalise them to dimensions 
different than four, a little bit of more work needs to be done.

\section{The Boundary Term and New Recursion Relations}\label{sec:Bound}

Let us observe a crucial fact. The one-parameter family of amplitudes $M_{n}^{\mbox{\tiny $(i,j)$}}(z)$
can be written as 
\begin{equation}\eqlabel{AmplZ}
 M_{n}^{\mbox{\tiny $(i,j)$}}(z)\:=\:\sum_{k\in\mathcal{P}^{\mbox{\tiny $(i,j)$}}}
 \frac{M_{\mbox{\tiny L}}^{\mbox{\tiny $(i,j)$}}(z_k)M_{\mbox{\tiny R}}^{\mbox{\tiny $(i,j)$}}(z_k)}{P^2_k(z)}
 +\mathcal{C}_{n}^{\mbox{\tiny $(i,j)$}}(z).
\end{equation}
Since all the singularities located at finite points are contained in the first sum in \eqref{AmplZ}, the
function $\mathcal{C}_{n}^{\mbox{\tiny $(i,j)$}}(z)$ is just a polynomial in $z$ of order $\nu$:
\begin{equation}\eqlabel{Cz}
 \mathcal{C}_{n}^{\mbox{\tiny $(i,j)$}}(z)\:=\:\mathcal{C}_{n}^{\mbox{\tiny $(i,j)$}}+
  \sum_{l=1}^{\nu}a_{l}^{\mbox{\tiny $(i,j)$}}z^l,
\end{equation}
where $\nu$ is as well the order the amplitude $M_n^{\mbox{\tiny $(i,j)$}}(z)$ diverges with, as $z$ is taken to infinity.
The zero-th order term in \eqref{Cz} is the only one which survives in the integral \eqref{AmplInt} and
therefore it is the only contribution to the physical amplitude. Here we are referring to a BCFW-deformation
defined as a shift of the momenta of two particles. This is however more general and holds for any type of
one-complex parameter deformation.

In order to understand the singularity at infinity, it is instructive to consider the logarithmic derivative 
of $M_{n}^{\mbox{\tiny $(i,j)$}}(z)$. Its pole structure is just given by simple poles at the location of both 
the zeroes and the poles of $M_{n}^{\mbox{\tiny $(i,j)$}}(z)$. Furthermore, the residues of such poles correspond
to the multiplicity of the zeroes and of the poles of $M_{n}^{\mbox{\tiny $(i,j)$}}(z)$. The BCFW-deformations induce 
just simple poles at finite locations and, therefore, the residues of the logarithmic derivative with respect to
such points is always $-1$. Taking into account the eventual multiplicity of the zeroes, the integration of the 
logarithmic derivative of $M_{n}^{\mbox{\tiny $(i,j)$}}(z)$ on the Riemann sphere provides with the following 
equation:
\begin{equation}\eqlabel{ArgPrinc}
 0\:=\:\int_{\mathcal{R}}dz\,\frac{M_{n}'^{\mbox{\tiny $(i,j)$}}(z)}{M_{n}^{\mbox{\tiny $(i,j)$}}(z)}\:=\:
  N_{\mbox{\tiny $Z$}}-N_{\mbox{\tiny $P$}}\,\equiv\,
  N_{\mbox{\tiny $Z$}}^{\mbox{\tiny fin}}-N_{\mbox{\tiny $P$}}^{\mbox{\tiny fin}}-\nu,
\end{equation}
which is just the generalisation of the argument principle to the Riemann sphere. In \eqref{ArgPrinc},
$N_{\mbox{\tiny $Z$}}$, $N_{\mbox{\tiny $P$}}$, $N_{\mbox{\tiny $Z$}}^{\mbox{\tiny fin}}$, 
$N_{\mbox{\tiny $P$}}^{\mbox{\tiny fin}}$ and $\nu$ are respectively the number of zeroes (with their
multiplicity), the number of poles, the number of zeroes at finite location, the number of poles at finite
location, and the multiplicity of the point at infinity. Equation \eqref{ArgPrinc} relates the latter to
$N_{\mbox{\tiny $Z$}}^{\mbox{\tiny fin}}$ and $N_{\mbox{\tiny $P$}}^{\mbox{\tiny fin}}$: 
$\nu\,=\,N_{\mbox{\tiny $Z$}}^{\mbox{\tiny fin}}-N_{\mbox{\tiny $P$}}^{\mbox{\tiny fin}}$. The number of poles 
at finite location $N_{\mbox{\tiny $P$}}^{\mbox{\tiny fin}}$ is known for a given theory. Thus, the knowledge
of the number of zeroes would fix the large-$z$ behaviour of the amplitudes from first principles. 

Let now $\left\{z_0^{\mbox{\tiny $(s)$}}\right\}$ be a subset of $n_{\mbox{\tiny $z$}}$ zeroes of the amplitude 
\eqref{AmplZ} and $\gamma_0^{\mbox{\tiny $(s)$}}$ be a contour including just $z_0^{\mbox{\tiny $(s)$}}$ and no
other zeroes or poles. Using the general expressions \eqref{AmplZ} and \eqref{Cz} for the one-parameter family 
of amplitudes $M_{n}^{\mbox{\tiny $(i,j)$}}(z)$, one obtains the following equations
\begin{equation}\eqlabel{zeroes}
 \begin{split}
  0\:=\:\frac{1}{2\pi i}\oint_{\mbox{\tiny $\gamma_0^{(s)}$}}dz\:
  &\frac{M_{n}^{\mbox{\tiny $(i,j)$}}(z)}{\left(z-z_0^{\mbox{\tiny $(s)$}}\right)^{r}}\:=\:
  \left(-1\right)^{r-1}\sum_{k=1}^{N_{\mbox{\tiny $P$}}^{\mbox{\tiny fin}}}
  \frac{M_{\mbox{\tiny L}}^{\mbox{\tiny $(i,j)$}}(z_k)M_{\mbox{\tiny R}}^{\mbox{\tiny $(i,j)$}}(z_k)}{
  \left(-2P_k\cdot q\right)\left(z_0^{\mbox{\tiny $(s)$}}-z_k\right)^{r}}+
  \delta_{\mbox{\tiny $r,1$}}\mathcal{C}_{\mbox{\tiny $n$}}^{\mbox{\tiny $(i,j)$}}+\\
  &\hspace{.2cm}+
  \sum_{l=1}^{\nu}\frac{l!}{\left(l-r+1\right)!\left(r-1\right)!}a_{l}^{\mbox{\tiny $(i,j)$}}z_{\mbox{\tiny $0$}}^{l-r+1},
  \quad \mbox{with }
  \left\{
   \begin{array}{l}
    r\,=\,1,\ldots,m^{\mbox{\tiny $(s)$}}\\
    s\,=\,1,\ldots,n_z
   \end{array}
  \right. ,
 \end{split}
\end{equation}
where $m^{\mbox{\tiny $(s)$}}$ is the multiplicity of the zero $z_0^{\mbox{\tiny $(s)$}}$. If $n_z\,=\,\nu+1$,
the system of algebraic equations \eqref{zeroes} would fix univocally the amplitudes. In particular, the boundary term 
$\mathcal{C}_{\mbox{\tiny $n$}}^{\mbox{\tiny $\left(i,j\right)$}}$ acquires the form\footnote{In \eqref{Cnij}, the product 
in square-brackets and the sums on the indices $\{r_k\}$ implicitly take into account the possibility of having zeroes with
multiplicity higher than one.}
\begin{equation}\eqlabel{Cnij}
 \begin{split}
  \mathcal{C}_{\mbox{\tiny $n$}}^{\mbox{\tiny $\left(i,j\right)$}}\:&=\:
   \left(-1\right)^{\nu+1}
   \sum_{k\in\mathcal{P}^{\mbox{\tiny $(i,j)$}}}
   \frac{M_{\mbox{\tiny L}}^{\mbox{\tiny $(i,j)$}}(z_k)M_{\mbox{\tiny R}}^{\mbox{\tiny $(i,j)$}}(z_k)}{P_k^2}
   \left[
    \prod_{l=1}^{\nu+1}\frac{P_k^2}{P_k^2(z_0^{\mbox{\tiny $(l)$}})}
   \right]\times\\
  &\times
   \sum_{l=0}^{\nu}\left(-1\right)^l
   \sum_{\mbox{\tiny $\begin{array}{l} r_1,..,r_l\,=\,1 \\ r_1\neq..\neq r_l \end{array}$}}^{\nu+1}
   \prod_{s=1}^{l}\frac{P_k^2\left(z_0^{\mbox{\tiny $(r_s)$}}\right)}{P_k^2}.
 \end{split}
\end{equation}
The expression \eqref{Cnij} determines the boundary term $\mathcal{C}_{\mbox{\tiny $n$}}^{\mbox{\tiny $\left(i,j\right)$}}$
in terms of the lower point on-shell amplitudes! This implies that the presence of the boundary term does not spoil
the standard BCFW structure! More precisely, the whole $n$-particle amplitude takes the form
\begin{equation}\eqlabel{Mfin}
 M_n\:=\:\sum_{k\in\mathcal{P}^{\mbox{\tiny $(i,j)$}}}
 \left[\frac{M_{\mbox{\tiny L}}^{\mbox{\tiny $(i,j)$}}(\hat{i},\mathcal{I}_k,-\hat{P}_{\mbox{\tiny $i\mathcal{I}_k$}})
  M_{\mbox{\tiny R}}^{\mbox{\tiny $(i,j)$}}(\hat{P}_{\mbox{\tiny $i\mathcal{I}_k$}}, \mathcal{J}_{k}, \hat{j})}{
  P_{\mbox{\tiny $i\mathcal{I}_k$}}^2}
 \prod_{l=1}^{\nu+1}\left(1-\frac{P_{\mbox{\tiny $i\mathcal{I}_k$}}^2}{P_{\mbox{\tiny $i\mathcal{I}_k$}}^2
  \left(z_0^{\mbox{\tiny $(l)$}}\right)}\right)\right],
\end{equation}
where $\mathcal{I}_{k}$ and $\mathcal{J}_{k}$ are sets of particles such that we have the partition
$\left\{i\right\}\cup\left\{j\right\}\cup\mathcal{I}_{k}\cup\mathcal{J}_{k}\,=\,\left\{1,\ldots,n\right\}$, which can 
contain from $1$ to $n-3$ elements, and the ``hat'' denotes that the momenta are computed at the location
of the pole.

Some comments are in order. As emphasised earlier, even in the presence of the boundary term, a scattering amplitude
can be expressed in terms of products of two on-shell scattering amplitudes with fewer external states. The statement that
the three-particle amplitudes determine the whole tree level is therefore generalized to any consistent theory of massless
particles with a non-trivial S-matrix! More precisely, the amplitudes are determined in terms of the smallest amplitudes 
defining a particular theory, which may have more external particles than three. However, it is always possible to define 
effective three-particle amplitudes for such theories, as it was done for $\lambda\phi^4$ in \cite{Benincasa:2007xk}. Even 
if this might not be technically convenient for the actual computation of an amplitude, it anyway points out
that the tree-level structure of these theories can be treated on the same footing as the ones for which the 
three-particle amplitudes are naturally defined.

Furthermore, the number of BCFW channels does not change, and the recursion relation
\eqref{Mfin} can be seen as a ``weighted'' version of the standard BCFW expression, and the ``weight'' is given by
the term dependent on the location of a subset of zeroes.

Finally, let us point out that the knowledge of a subset of $\nu+1$ zeroes allows us to determine the full structure
of the one-parameter family of amplitudes defined by a given BCFW-deformation. In particular, we can explicitly determine
as well the leading order in $z$, as $z$ is taken to infinity, which has the interpretation of a hard particle 
in a soft background \cite{ArkaniHamed:2008yf}
\begin{equation}\eqlabel{HardP}
 a_{\nu}^{\left(i,j\right)}\:=\:(-1)^{\nu+1}\sum_{k\in\mathcal{P}^{\mbox{\tiny $(i,j)$}}}
  \frac{M_{\mbox{\tiny L}}^{\mbox{\tiny $(i,j)$}}(\hat{i},\mathcal{I}_k,-\hat{P}_{\mbox{\tiny $i\mathcal{I}_k$}})
  M_{\mbox{\tiny R}}^{\mbox{\tiny $(i,j)$}}(\hat{P}_{\mbox{\tiny $i\mathcal{I}_k$}}, \mathcal{J}_{k}, \hat{j})}{
  2\,P_{\mbox{\tiny $i\mathcal{I}_k$}}\cdot q}
 \prod_{l=1}^{\nu+1}\frac{2\,P_{\mbox{\tiny $i\mathcal{I}_k$}}\cdot q}{P_{\mbox{\tiny $i\mathcal{I}_k$}}^{2}
  \left(z_0^{\mbox{\tiny $(l)$}}\right)}.
\end{equation}
Notice that for $\nu\,=\,0$, {\it i.e.} $M_n^{\mbox{\tiny $(i,j)$}}(z)\,\overset{z\rightarrow\infty}{\sim}
 \mathcal{O}\left(1\right)$, the expressions \eqref{Cnij} for the boundary term and \eqref{HardP} for the
hard-particle limit coincide, as it should be.

\section{Zeroes, UV and Collinear/Multi-particle Limits}\label{sec:ZeroLimits}

In the previous section, we discussed how the knowledge of a subset of zeroes can fix the large-$z$ behaviour of the
amplitude under a certain BCFW-deformation, which is given by the difference between the number of zeroes and the
number of poles, and the boundary term through their location. The result is the recursion relation \eqref{Mfin},
which shows the same number of terms as the standard BCFW formula.

The question we need to answer now concerns the physical meaning of the location of the zeroes of the amplitudes, or, 
probably more fruitfully, the physical meaning of the internal propagators when they are evaluated at the location
of the zeroes.

Beyond the obvious statement that the location of the zeroes are points in the complexified momentum space where
the S-matrix becomes trivial, to our knowledge not much is known about their physical significance. The issue of the
zeroes of the complete amplitudes was studied in relation to the analysis of the dispersion relations for the
logarithm of the scattering amplitudes 
\cite{Sugawara:1963aa, Jin:1964zz, Vernov:1975au, Zamiralov:1976qc, Kurbatov:1977xb, Vernov:1980xv}.

In the present case, the relation \eqref{Mfin} indeed provides us with a new representation of the full $n$-particle 
amplitudes for arbitrary theories, and therefore it must provide us with the correct collinear/multi-particle limits. 
We will show how the zeroes of the amplitudes and its large-$z$ behaviour are fixed by such limits. The
collinear/multi-particle channels we will need to analyse can be grouped in the following four classes:
\begin{equation}\eqlabel{colmul}
 \begin{split}
  &\lim_{\mbox{\tiny $P_{i\mathcal{I}_k}^2$}\:\rightarrow\:0}
    P_{\mbox{\tiny $i\mathcal{I}_k$}}^2\,M_n\:=\:
   M(i,\,\mathcal{I}_k,\,-P_{i\mathcal{I}_k})\, M(P_{i\mathcal{I}_k}, \mathcal{J}_k, j),\\
  &\lim_{\mbox{\tiny $P_{\mathcal{K}}^2$}\:\rightarrow\:0}
    P_{\mbox{\tiny $\mathcal{K}$}}^2\,M_n\:=\:
   M_{s+1}(\mathcal{K},\,-P_{\mathcal{K}})\, M_{n-s+1}(P_{\mathcal{K}}, \mathcal{Q}, i, j),\\
  &\lim_{\mbox{\tiny $P_{k_1 k_2}^2$}\:\rightarrow\:0}
    P_{\mbox{\tiny $k_1 k_2$}}^2\,M_n\:=\:
   M_3(k_1,\,k_2,\,-P_{k_1 k_2})\, M_{n-1}(P_{k_1 k_2}, \mathcal{K}, i, j),\\
  &\lim_{\mbox{\tiny $P_{ij}^2$}\:\rightarrow\:0}
    P_{\mbox{\tiny $ij$}}^2\,M_n\:=\:
   M_{3}(i,\,j,\,-P_{ij})\, M_{n-1}(P_{ij}, \mathcal{K}),\\
 \end{split}
\end{equation}
where $\mathcal{K}$ in the second line indicates a set of $s\,\ge\,3$ particles and $\mathcal{Q}$ a set of $n-s-2$ 
particles, so that, together with $\{i\}$ and $\{j\}$, they form the partition 
$\mathcal{K}\,\cup\,\mathcal{Q}\,\cup\,\left\{i\right\}\,\cup\,\left\{j\right\}\,=\,\left\{1,\ldots,n\right\}$.

Before starting with the discussion of the limits \eqref{colmul}, it is useful to write down here, for future
reference, all the quantities characterising our recursive formula \eqref{Mfin}. As mentioned earlier, the recursive
structure \eqref{Mfin} has been obtained under the deformation \eqref{BCFWdef}. The induced one-parameter family of
amplitudes $M_{\mbox{\tiny $n$}}^{\mbox{\tiny $(i,j)$}}(z)$ shows poles located at
\begin{equation}\eqlabel{poles2}
 z_{\mbox{\tiny $i\mathcal{I}_k$}}\:=\:\frac{P_{\mbox{\tiny $i\mathcal{I}_k$}}^2}{2
  P_{\mbox{\tiny $i\mathcal{I}_k$}}\cdot q}.
\end{equation}
The amplitudes with fewer external states contained in \eqref{Mfin} are computed at the locations \eqref{poles2}, where
the deformed momenta have the explicit form
\begin{equation}\eqlabel{defmom}
 \begin{split}
  &\hat{p}^{\mbox{\tiny $(i)$}}\:\overset{\mbox{\tiny def}}{=}\:p^{\mbox{\tiny $(i)$}}
    \left(z_{\mbox{\tiny $i\mathcal{I}_k$}}\right)\:=\:
   p^{\mbox{\tiny $(i)$}}-\frac{P_{\mbox{\tiny $i\mathcal{I}_k$}}^2}{2 P_{\mbox{\tiny $i\mathcal{I}_k$}}\cdot q}q,
   \quad
   \hat{p}^{\mbox{\tiny $(j)$}}\:\overset{\mbox{\tiny def}}{=}\:p^{\mbox{\tiny $(j)$}}
    \left(z_{\mbox{\tiny $i\mathcal{I}_k$}}\right)\:=\:
   p^{\mbox{\tiny $(j)$}}+\frac{P_{\mbox{\tiny $i\mathcal{I}_k$}}^2}{2 P_{\mbox{\tiny $i\mathcal{I}_k$}}\cdot q}q,\\
  &\hat{P}_{\mbox{\tiny $i\mathcal{I}_k$}}\:\overset{\mbox{\tiny def}}{=}\:P_{\mbox{\tiny $i\mathcal{I}_k$}}
    \left(z_{\mbox{\tiny $i\mathcal{I}_k$}}\right)\:=\:
   P_{\mbox{\tiny $i\mathcal{I}_k$}}-
   \frac{P_{\mbox{\tiny $i\mathcal{I}_k$}}^2}{2 P_{\mbox{\tiny $i\mathcal{I}_k$}}\cdot q}q.
 \end{split}
\end{equation}

One further comment is in order. Let us explicitly consider the four-dimensional helicity-spinor formalism. In a generic
$(k_1,\,k_2)$-channel, the limit $P_{\mbox{\tiny $k_1 k_2$}}^2\,\rightarrow\,0$ can be taken in two different ways. 
Specifically, $P_{\mbox{\tiny $k_1 k_2$}}^2\,=\,\langle k_1,\,k_2\rangle[k_1,\,k_2]$ with the two Lorentz invariant
inner products independent of each other which can therefore be taken to zero in the complexified momentum space, as 
either $\langle k_1,\,k_2\rangle$ or $[k_1,\,k_2]$ go to zero. This will be of crucial importance in the analysis of
the collinear limits of \eqref{Mfin} and it will be further discussed later.

\subsection{The collinear/multi-particle limit $P_{i\mathcal{I}_k}^2\,\rightarrow\,0$}

This limit is manifestly satisfied by the set of recursion relations \eqref{Mfin} we propose. However, for the sake
of completeness, we analyse it in detail. From \eqref{defmom}, it is easy to see that in this kinematic limit
\begin{equation}\eqlabel{lim1}
 \hat{p}^{\mbox{\tiny $(i)$}}\:\rightarrow\:p^{\mbox{\tiny $(i)$}}, \quad
 \hat{p}^{\mbox{\tiny $(j)$}}\:\rightarrow\:p^{\mbox{\tiny $(j)$}}, \quad
 \hat{P}_{\mbox{\tiny $i\mathcal{I}_k$}}\:\rightarrow\:P_{\mbox{\tiny $i\mathcal{I}_k$}},
\end{equation}
so that
\begin{equation}\eqlabel{lim1b}
  \lim_{\mbox{\tiny $P_{i\mathcal{I}_k}^2$}\:\rightarrow\:0}P_{\mbox{\tiny $i\mathcal{I}_k$}}^2\,M_n\:=\:
  M_{\mbox{\tiny L}}^{\mbox{\tiny $(i,j)$}}(\hat{i},\mathcal{I}_k,-\hat{P}_{\mbox{\tiny $i\mathcal{I}_k$}})
  M_{\mbox{\tiny R}}^{\mbox{\tiny $(i,j)$}}(\hat{P}_{\mbox{\tiny $i\mathcal{I}_k$}}, \mathcal{J}_{k}, \hat{j}),
\end{equation}
and
\begin{equation}\eqlabel{lim1b2}
  \lim_{\mbox{\tiny $P_{i\mathcal{I}_k}^2$}\:\rightarrow\:0}f_{i\mathcal{I}_k}^{\mbox{\tiny $(\nu,n)$}}\,=\,1.
\end{equation}
Indeed this is not a surprise given that the deformation \eqref{BCFWdef} selects all the factorisation channels
$P_{i\mathcal{I}_k}^2\,\rightarrow\,0$ (figure \ref{fig:wBCFW}). 

As emphasised in \cite{Schuster:2008nh}, in the case where either $\mathcal{I}_k$ or $\mathcal{J}_{k}$ contains just
one particle, the relevant collinear limit is related to only one of the Lorentz invariant spinorial inner products.
Specifically, under the deformation \eqref{BCFWdef2}, for $\mathcal{I}_k\,=\,\left\{k\right\}$ the relevant singularity is 
provided by $[i,\,k]\,\longrightarrow\,0$, while for $\mathcal{J}_k\,=\,\left\{k\right\}$ by 
$\langle j,\,k\rangle\,\longrightarrow0$.

\subsection{The multi-particle limit $P_{\mathcal{K}}^2\,\rightarrow\,0$}\label{subsec:PK}

We now consider the factorisation along a class of channels $\mathcal{K}$ which contain at least three particles:
$P_{\mathcal{K}}^2\,\rightarrow\,0$. In the recursive formula \eqref{Mfin}, let us take $\mathcal{K}$ to be
$\mathcal{K}\,=\,\mathcal{K}_{\mathcal{I}}\,\cup\,\mathcal{K}_{\mathcal{J}}$, with 
$\mathcal{K}_{\mathcal{I}}\,\subseteq\,\mathcal{I}_k$ and $\mathcal{K}_{\mathcal{J}}\,\subseteq\,\mathcal{J}_k$
- we can think about $\mathcal{I}_k$ and $\mathcal{J}_k$ as 
$\mathcal{I}_k\,=\,\mathcal{K}_{\mbox{\tiny $\mathcal{I}$}}\,\cup\,\tilde{\mathcal{I}}_k$ and
$\mathcal{J}_k\,=\,\mathcal{K}_{\mbox{\tiny $\mathcal{J}$}}\,\cup\,\tilde{\mathcal{J}}_k$.
The sum in \eqref{Mfin} can be split into three classes of terms:
$\left(\mathcal{K}_{\mathcal{I}},\,\mathcal{K}_{\mathcal{J}}\right)\,=\,\left\{\left(\mathcal{K},\,\emptyset\right),\,
  \left(\emptyset,\,\mathcal{K}\right),\,\left(\mathcal{K}_{\mathcal{I}},\,\mathcal{K}_{\mathcal{J}}\right)\right\}$.

Taking the limit $\mathcal{P}_{\mathcal{K}}^2\rightarrow0$, 
the set of on-shell diagrams 
$\left(\mathcal{K}_{\mathcal{I}},\,\mathcal{K}_{\mathcal{J}}\right)\,=\,\left(\mathcal{K},\,\emptyset\right)$ 
shows the singularity $P_{\mathcal{K}}^2$ in the sub-amplitude
$M_{\mbox{\tiny L}}^{\mbox{\tiny $(i,j)$}}(\hat{i},\mathcal{I}_k,-\hat{P}_{\mbox{\tiny $i\mathcal{I}_k$}})$,
which factorises as follows
\begin{equation}\eqlabel{lim2a}
 M_{\mbox{\tiny L}}^{\mbox{\tiny $(i,j)$}}(\hat{i},\mathcal{I}_k,-\hat{P}_{\mbox{\tiny $i\mathcal{I}_k$}})
 \:\overset{\mbox{\tiny $P_{\mathcal{K}}^2\,\rightarrow\,0$}}{\longrightarrow}\:
 M_{s+1}(\mathcal{K},\,-P_{\mathcal{K}})\frac{1}{P_{\mathcal{K}}^2}
 M_{\mbox{\tiny L}}^{\mbox{\tiny $(i,j)$}}
 (P_{\mathcal{K}},\hat{i},\tilde{\mathcal{I}}_{k},-\hat{P}_{\mbox{\tiny $i\mathcal{I}_k$}}).
\end{equation}
In the second case,
$\left(\mathcal{K}_{\mathcal{I}},\,\mathcal{K}_{\mathcal{J}}\right)\,=\,\left(\emptyset,\,\mathcal{K}\right)$,
the same type of factorisation occurs in the sub-amplitude
$M_{\mbox{\tiny R}}^{\mbox{\tiny $(i,j)$}}(\hat{P}_{\mbox{\tiny $i\mathcal{I}_k$}}, \mathcal{J}_{k}, \hat{j})$:
\begin{equation}\eqlabel{lim2b}
 M_{\mbox{\tiny R}}^{\mbox{\tiny $(i,j)$}}(\hat{P}_{\mbox{\tiny $i\mathcal{I}_k$}}, \mathcal{J}_{k}, \hat{j})
 \:\overset{\mbox{\tiny $P_{\mathcal{K}}^2\,\rightarrow\,0$}}{\longrightarrow}\:
 M_{s+1}(\mathcal{K},\,-P_{\mathcal{K}})\frac{1}{P_{\mathcal{K}}^2}
 M_{\mbox{\tiny R}}^{\mbox{\tiny $(i,j)$}}
 (P_{\mathcal{K}},\hat{P}_{\mbox{\tiny $i\mathcal{I}_k$}},\tilde{\mathcal{J}}_{k},\hat{j})
\end{equation}
Finally, in the third class of terms neither of the two sub-amplitudes show a singularity in this channel. 
One might think that the ``weight'' may show such singularity.
However, this cannot be the case given that this condition will not induce this class of terms to factorise and 
generate a sub-amplitude of type $M_{s+1}(\mathcal{K},\,-P_{\mathcal{K}})$: this would spoil the possibility to recover
the correct factorisation properties along this class of channels.

\begin{figure}[htbp]
 \centering 
  \scalebox{.35}{\includegraphics{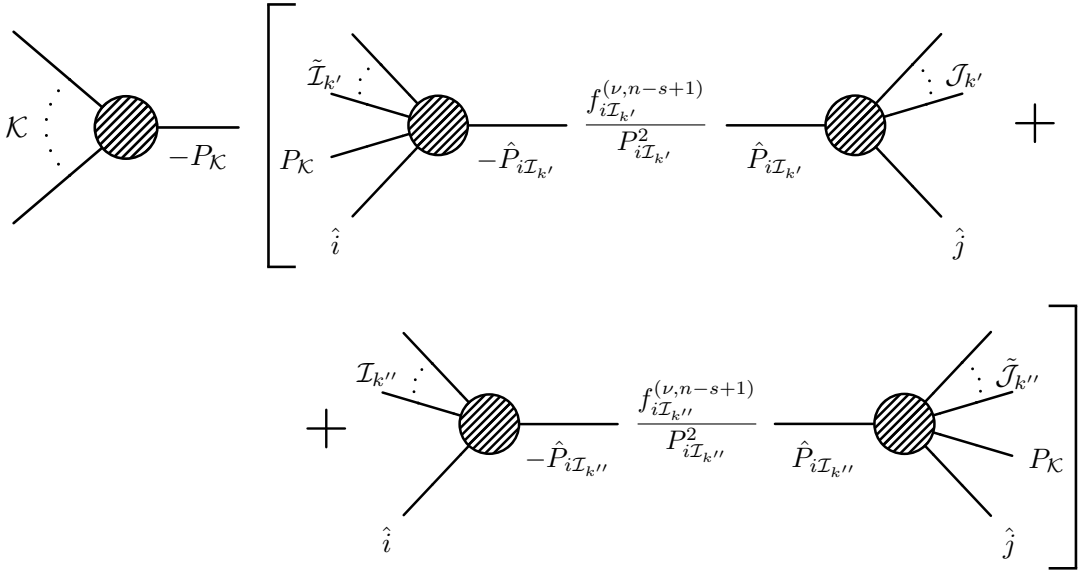}}
  \caption{Multi-particle limit $P_{\mathcal{K}}^2\,\rightarrow\,0$. In this figure, the behaviour of the terms of the
           generalised recursion relation \eqref{Mfin} under the multi-particle
           limits is depicted. In particular, in the terms contributing, all the particles in the set $\mathcal{K}$ belong
           to one of the two sub-amplitudes.}\label{fig:PK}
\end{figure}

Therefore, \eqref{Mfin} in the limit $\mathcal{P}_{\mathcal{K}}^2\,\rightarrow\,0$ becomes
\begin{equation}\eqlabel{lim2c}
 \begin{split}
 \lim_{\mathcal{P}_{\mathcal{K}}^2\,\rightarrow\,0}&\mathcal{P}_{\mathcal{K}}^2\,M_n\:=\:
  M_{s+1}(\mathcal{K},\,-P_{\mathcal{K}})\times\\
 &\times
  \left[
   \sum_{k'\in\mathcal{P}^{\mbox{\tiny $(i,j)$}}}
    \frac{M_{\mbox{\tiny L}}^{\mbox{\tiny $(i,j)$}}
     (P_{\mathcal{K}},\hat{i},\tilde{\mathcal{I}}_{k'},-\hat{P}_{\mbox{\tiny $i\mathcal{I}_{k'}$}})
     M_{\mbox{\tiny R}}^{\mbox{\tiny $(i,j)$}}(\hat{P}_{\mbox{\tiny $i\mathcal{I}_{k'}$}}, \mathcal{J}_{k'}, \hat{j})}{
     P_{\mbox{\tiny $i\mathcal{I}_k'$}}^2}
     \prod_{l=1}^{\nu+1}\left(1-\frac{P_{\mbox{\tiny $i\mathcal{I}_{k'}$}}^2}{
     P_{\mbox{\tiny $i\mathcal{I}_{k'}$}}^2\left(\bar{z}_0^{\mbox{\tiny $(l)$}}\right)}\right)+
   \right.\\
  &\left.+
   \sum_{k''\in\mathcal{P}^{\mbox{\tiny $(i,j)$}}}
   \frac{M_{\mbox{\tiny L}}^{\mbox{\tiny $(i,j)$}}(\hat{i},\mathcal{I}_{k''},-\hat{P}_{\mbox{\tiny $i\mathcal{I}_{k''}$}})
    M_{\mbox{\tiny R}}^{\mbox{\tiny $(i,j)$}}
   (P_{\mathcal{K}},\hat{P}_{\mbox{\tiny $i\mathcal{I}_{k''}$}},\tilde{\mathcal{J}}_{k''},\hat{j})}{
   P_{\mbox{\tiny $i\mathcal{I}_{k''}$}}^2}
    \prod_{l=1}^{\nu+1}\left(1-\frac{P_{\mbox{\tiny $i\mathcal{I}_{k''}$}}^2}{
     P_{\mbox{\tiny $i\mathcal{I}_{k''}$}}^2\left(\bar{z}_0^{\mbox{\tiny $(l)$}}\right)}\right)
  \right],
 \end{split}
\end{equation}
where $\bar{z}_{0}^{\mbox{\tiny $(l)$}}$ indicates the zero $z_{0}^{\mbox{\tiny $(l)$}}$ in the limit 
$P_{\mbox{\tiny $\mathcal{K}$}}^2\,\rightarrow\,0$, as depicted in figure \ref{fig:PK}

Notice that the expression in the square brackets in \eqref{lim2c} is nothing but our formula \eqref{Mfin} for an amplitude
with fewer external legs! Specifically, if $s\,\ge\,3$ is the number of elements of $\mathcal{K}$, then the term in square
brackets in \eqref{lim2c} represents an on-shell recursion relation for an $(n-s+1)$-particle amplitude. There is one 
striking feature emerging from \eqref{lim2c}:
the large-$z$ (complex-UV) behaviour $\nu$ appearing for the $(n-s+1)$-particle amplitudes is the same as the one for 
$n$-particles! 

One might wonder whether some or all of the terms in the product defining the weights 
$f_{\mbox{\tiny $i\mathcal{I}_k$}}^{\mbox{\tiny $(\nu,n)$}}$ of the $n$-point amplitude goes to one, so that
the power with which the $(n-s+1)$-particle amplitudes diverges as $z$ is taken to infinity changes.
This is indeed not the case and it will be manifest from the analysis of the collinear limit $P_{ij}^2\,\rightarrow\,0$ 
(section \ref{subsec:Pij}).

One confusion that might arise from \eqref{lim2c} is that it might seem to state that the locations of the zeroes for the 
$n$-particle amplitude and for the $(n-s+1)$-particle amplitudes would coincide. Indeed this should not be the case and it is
not implied by \eqref{lim2c}. It simply states that the location of some zeroes of the $(n-s+1)$-particle amplitudes can
be obtained as a limit of the one for the zeroes of the $n$-particle amplitude when $P_{\mathcal{K}}^2\,\rightarrow\,0$:
\begin{equation}\eqlabel{z0K}
 \lim_{P_{\mathcal{K}}^2\,\rightarrow\,0}P_{\mbox{\tiny $i\mathcal{I}_k$}}^2(z_0^{\mbox{\tiny $(l)$}})\:=\:
 P_{\mbox{\tiny $i\mathcal{I}_k$}}^2(\bar{z}_0^{\mbox{\tiny $(l)$}}),
\end{equation}
where $\bar{z}_0^{\mbox{\tiny $(l)$}}$ is a zero of the $(n-s+1)$-particle amplitude 
$M_{n-s+1}(i,\,j,\,\mathcal{Q},\,P_{\mbox{\tiny $\mathcal{K}$}})$.

Thinking about the large-$z$ limit as a hard particle of momentum $zq$ propagating in a soft background, this is the 
statement that the soft particles do not affect the leading (complex) UV behaviour.

\subsection{The collinear limit $P_{\mbox{\tiny $k_1 k_2$}}^2\,\rightarrow\,0$}\label{subsec:Pk1k2}

In the previous section, we discussed the multi-particle limit $P_{\mathcal{K}}^2\,\rightarrow\,0$, with $\mathcal{K}$
containing $s\,\ge\,3$ particles. Here we turn to the two-particle version of this limit, which is more subtle, as
pointed out in \cite{Schuster:2008nh} for gluons and gravitons. However, before starting with the analysis of this collinear
limit on the recursion relation \eqref{Mfin}, it is useful to comment on the expected result (third line of equation 
\eqref{colmul}):
\begin{equation}\eqlabel{Pk1k2lim}
 \lim_{\mbox{\tiny $P_{k_1 k_2}^2$}\:\rightarrow\:0}
  P_{\mbox{\tiny $k_1 k_2$}}^2\,M_n\:=\: M_3(k_1,\,k_2,\,-P_{k_1 k_2})\, M_{n-1}(P_{k_1 k_2}, \mathcal{K}, i, j).
\end{equation}
As mentioned earlier, there are two possible ways in which the limit $P_{\mbox{\tiny $k_1 k_2$}}^2\,\rightarrow\,0$ can be 
taken. Specifically, considering that $P_{\mbox{\tiny $k_1 k_2$}}^2\,=\,\langle k_1,\,k_2\rangle\,[k_1,\,k_2]$ and that
the spinors related to the same momentum are independent of each other in the complexified momentum space, it is
possible to consider $\langle k_1,\,k_2\rangle\,\rightarrow\,0$ or $[k_1,\,k_2]\,\rightarrow\,0$. In the case,
$\langle k_1,\,k_2\rangle\,\rightarrow\,0$, the holomorphic spinors in the three-particle amplitude on the right-hand-side 
of \eqref{Pk1k2lim} are proportional to each other, so that $M_3(k_1,\,k_2,\,-P_{k_1 k_2})$ is just a function of the
 anti-holomorphic spinors \cite{Benincasa:2007xk} for theories with three-particle $\delta$-derivative interactions, with 
$\delta\,>\,0$. Similarly, when the anti-holomorphic inner product $[k_1,\,k_2]$ is taken to zero, the three-particle 
amplitude is just function of the holomorphic spinors (again, for theories with three-particle $\delta$-derivative 
interactions, with $\delta\,>\,0$).

Let us analyze the expected factorisation \eqref{Pk1k2lim} in both complex collinear limits, starting with theories
with three-particle $\delta$-derivative interactions ($\delta\,>\,0$). As 
$\langle k_1,\,k_2\rangle\,\rightarrow\,0$, $\lambda^{\mbox{\tiny $(k_1)$}}\,\sim\,\lambda^{\mbox{\tiny $(k_2)$}}$
and the momentum $P_{k_1 k_2}$ may acquire the form
\begin{equation}\eqlabel{Pk1k2hol}
 \begin{split}
  P_{k_1 k_2}\:\overset{\mbox{\tiny def}}{=}\:\lambda^{\mbox{\tiny $(k_1)$}}\tilde{\lambda}^{\mbox{\tiny $(k_1)$}}+
  \lambda^{\mbox{\tiny $(k_2)$}}\tilde{\lambda}^{\mbox{\tiny $(k_2)$}}\:
  &\overset{\mbox{\tiny $\langle k_1,\,k_2\rangle\,\rightarrow\,0$}}{\longrightarrow}
  \lambda^{\mbox{\tiny $(k_1)$}}\left(\tilde{\lambda}^{\mbox{\tiny $(k_1)$}}+
  \frac{\langle k_2,\,\mu\rangle}{\langle k_1,\,\mu\rangle}\tilde{\lambda}^{\mbox{\tiny $(k_2)$}}\right),\\
  &\overset{\mbox{\tiny $\langle k_1,\,k_2\rangle\,\rightarrow\,0$}}{\longrightarrow}
   \lambda^{\mbox{\tiny $(k_2)$}}\left(\frac{\langle k_1,\,\mu\rangle}{\langle k_2,\,\mu\rangle}
    \tilde{\lambda}^{\mbox{\tiny $(k_1)$}}+\tilde{\lambda}^{\mbox{\tiny $(k_2)$}}\right),
 \end{split}
\end{equation}
for some reference spinor $\mu$. The two expressions in \eqref{Pk1k2hol} depend on whether the holomorphic spinor
$\lambda^{\mbox{\tiny $(k_1 k_2)$}}$ of $P_{k_1 k_2}$ is chosen to be $\lambda^{\mbox{\tiny $(k_1)$}}$ or 
$\lambda^{\mbox{\tiny $(k_2)$}}$. One has indeed the freedom to relate $\lambda^{\mbox{\tiny $(k_1 k_2)$}}$
to $\lambda^{\mbox{\tiny $(k_1)$}}$ or $\lambda^{\mbox{\tiny $(k_2)$}}$ through proportionality factors different
from one, at the price of changing the other proportionality factors in front of the anti-holomorphic spinors
in \eqref{Pk1k2hol}. Let us choose the following identification
\begin{equation}\eqlabel{Pk1Pk2id}
 \lambda^{\mbox{\tiny $(k_1 k_2)$}}\:=\:\lambda^{\mbox{\tiny $(k_1)$}},\qquad
 \tilde{\lambda}^{\mbox{\tiny $(k_1 k_2)$}}\:=\:\tilde{\lambda}^{\mbox{\tiny $(k_1)$}}+
  \frac{\langle k_2,\,\mu\rangle}{\langle k_1,\,\mu\rangle}\tilde{\lambda}^{\mbox{\tiny $(k_2)$}}.
\end{equation}
The three-particle amplitude on the right-hand-side of \eqref{Pk1k2lim} becomes
\begin{equation}\eqlabel{Pk1k23pt}
 \begin{split}
  M_3&(k_1^{\mbox{\tiny $h_1$}},\,k_2^{\mbox{\tiny $h_2$}},\,-P_{k_1 k_2}^{\mbox{\tiny $h_{12}$}})\:=\\
   &=\:
   \kappa_{\mbox{\tiny $1-\delta$}}[k_1,\,k_2]^{\mbox{\tiny $-(h_{12}-h_1-h_2)$}}[k_2,\,P_{k_1 k_2}]^{-(h_1-h_2-h_{12})}
   [P_{k_1 k_2},\,k_1]^{-(h_2-h_{12}-h_1)}\:=\\
   &=\:\kappa_{\mbox{\tiny $1-\delta$}}(-1)^{2h_1+2h_2}[k_1,\,k_2]^{\delta}
    \left(\frac{\langle k_1,\,\mu\rangle}{\langle k_2,\,\mu\rangle}\right)^{2h_2-\delta},
 \end{split}
\end{equation}
where the subscript in the coupling constant indicates the dimensionality of the coupling constant itself and we used the
fact that the number of derivatives $\delta$ of the three-particle interaction is related to the helicities of the
particles in a anti-holomorphic three-particle amplitude through the relation $\delta\,=\,h_1+h_2+h_{12}$ 
\cite{Benincasa:2007xk}.

The case $[k_1,\,k_2]\,\rightarrow\,0$ is treated similarly, with the spinorial representation of $P_{k_1 k_2}$ 
obtained from \eqref{Pk1k2hol} and \eqref{Pk1Pk2id} by exchanging holomorphic and anti-holomorphic spinors, and the
three-particle amplitude acquires the form
\begin{equation}\eqlabel{Pk1k23pt2}
 \begin{split}
  M_3&(k_1^{\mbox{\tiny $h_1$}},\,k_2^{\mbox{\tiny $h_2$}},\,-P_{k_1 k_2}^{\mbox{\tiny $h_{12}$}})\:=\\
   &=\:
   \kappa_{\mbox{\tiny $1-\delta$}}\langle k_1,\,k_2\rangle^{\mbox{\tiny $h_{12}-h_1-h_2$}}
    \langle k_2,\,P_{k_1 k_2}\rangle^{h_1-h_2-h_{12}} \langle P_{k_1 k_2},\,k_1\rangle^{h_2-h_{12}-h_1}\:=\\
   &=\:\kappa_{\mbox{\tiny $1-\delta$}}(-1)^{2h_1+2h_2}\langle k_1,\,k_2\rangle^{\delta}
    \left(\frac{[k_2,\,\mu]}{[k_1,\,\mu]}\right)^{2h_2+\delta},
 \end{split}
\end{equation}
where we used the relation $\delta\,=\,-h_1-h_2-h_{12}$ which characterises the holomorphic three-particle amplitude.

Notice that if we restrict ourselves to a theory with $\delta$-derivative three-particle interactions only, as we are doing,
the helicity of $P_{k_1 k_2}$ is univocally fixed by the relations $\delta\,=\,-h_1-h_2-h_{12}$, if the three-particle
amplitude is function of the holomorphic spinors only, or $\delta\,=\,h_1+h_2+h_{12}$ if it is just a function of the
anti-holomorphic spinors.

In the case of $\delta\,=\,0$, the helicity of $P_{k_1 k_2}$ is fixed to be $h_{12}\,=\,-h_1-h_2$ and
the general form of
$M_3(k_1^{\mbox{\tiny $h_1$}},\,k_2^{\mbox{\tiny $h_2$}},\,-P_{k_1 k_2}^{\mbox{\tiny $h_{12}$}})$
is
\begin{equation}\eqlabel{Pk1k2d0}
 \begin{split}
 M_3&(k_1^{\mbox{\tiny $h_1$}},\,k_2^{\mbox{\tiny $h_2$}},\,-P_{k_1 k_2}^{\mbox{\tiny $h_{12}$}})\:=\\
 &=\:
 \kappa_{\mbox{\tiny H}}\frac{\langle k_2,\,P_{k_1 k_2}\rangle^{2h_1}\langle P_{k_1 k_2},\,k_1\rangle^{2h_2}}{
   \langle k_1,\,k_2\rangle^{2h_1+2h_2}}+
 \kappa_{\mbox{\tiny A}}\frac{[k_1,\,k_2]^{2h_1+2h_2}}{[k_2,\,P_{k_1 k_2}]^{2h_1}[P_{k_1 k_2},\,k_1]^{2h_2}}.
 \end{split}
\end{equation}
In the limits $\langle k_1,\,k_2\rangle\,\rightarrow\,0$ and $[k_1,\,k_2]\,\rightarrow\,0$, the above expression reads
\begin{equation}\eqlabel{Pk1k2d02}
 \begin{split}
  &\left.M_3(k_1^{\mbox{\tiny $h_1$}},\,k_2^{\mbox{\tiny $h_2$}},\,-P_{k_1 k_2}^{\mbox{\tiny $h_{12}$}})\right|_{
    \mbox{\tiny $\langle k_1,\,k_2\rangle\,\rightarrow\,0$}}\:=\:
   \hat{\kappa}(-1)^{2h_1+2h_2}\left(\frac{\langle k_1,\,\mu\rangle}{\langle k_2,\,\mu\rangle}\right)^{2h_2},\\
  &\left.M_3(k_1^{\mbox{\tiny $h_1$}},\,k_2^{\mbox{\tiny $h_2$}},\,-P_{k_1 k_2}^{\mbox{\tiny $h_{12}$}})\right|_{
    \mbox{\tiny $[k_1,\,k_2]\,\rightarrow\,0$}}\:=\:
    \hat{\kappa}(-1)^{2h_1+2h_2}\left(\frac{[k_2,\,\mu]}{[k_1,\,\mu]}\right)^{2h_2},
 \end{split}
\end{equation}
with $\hat{\kappa}\,=\,\kappa_{\mbox{\tiny H}}+\kappa_{\mbox{\tiny A}}$ being the effective three-particle coupling.

Now we are ready to analyse these complex collinear limits on the recursive relation \eqref{Mfin}. As in
section \ref{subsec:PK}, there are three classes of terms to analyse and we will do it in the next subsections.

\subsubsection{Particles $k_1$ and $k_2$ on the same sub-amplitude}\label{subsub:same}

First, let us focus on the case in which $k_1$ and $k_2$ belong to the same sub-amplitude and take the limit 
$[k_1,k_2]\,\rightarrow\,0$. The analysis follows the one in the 
section \ref{subsec:PK} as long as $\tilde{\mathcal{I}}_{k}$ and $\tilde{\mathcal{J}}_{k}$ are not empty. 
If $k_1,\,k_2\,\in\,\mathcal{I}_k$ and $\tilde{\mathcal{I}}_{k}\,=\,\emptyset$, the sub-amplitude 
$M_{\mbox{\tiny L}}^{\mbox{\tiny $(i,j)$}}(\hat{i},\mathcal{I}_k,-\hat{P}_{\mbox{\tiny $i\mathcal{I}_k$}})$ is
a four-particle amplitude with channels $P_{\hat{i}k_1}^2$, $P_{\hat{i}k_2}^2$ and $P_{k_1 k_2}^2$. 
It turns out that $\hat{\tilde{\lambda}}^{(i)} \propto \tilde{\lambda}^{(k_1)} \propto
\tilde{\lambda}^{(k_2)}$ and therefore all the channels do contribute. 
The amplitude $M_{\mbox{\tiny L}}^{\mbox{\tiny $(i,j)$}}(\hat{i},\mathcal{I}_k,-\hat{P}_{\mbox{\tiny $i\mathcal{I}_k$}})$
therefore factorises as:
\begin{equation}\eqlabel{lim3b}
 \begin{split}
  M_{\mbox{\tiny L}}^{\mbox{\tiny $(i,j)$}}(\hat{i},\mathcal{I}_k,-\hat{P}_{\mbox{\tiny $i\mathcal{I}_k$}})\:&\equiv\:
  M_{\mbox{\tiny L}}^{\mbox{\tiny $(i,j)$}}(\hat{i},k_1,k_2,-\hat{P}_{\mbox{\tiny $i k_1 k_2$}})
  \:\overset{\mbox{\tiny $P_{k_1 k_2}^2\,\rightarrow\,0$}}{\longrightarrow}\\
  &\overset{\mbox{\tiny $P_{k_1 k_1}^2\,\rightarrow\,0$}}{\longrightarrow}\:
   M_3(k_1,k_2,-P_{\mbox{\tiny $k_1 k_2$}})\frac{1}{P_{\mbox{\tiny $k_1 k_2$}}^2}
   M_{\mbox{\tiny L}}^{\mbox{\tiny $(i,j)$}}(P_{\mbox{\tiny $k_1 k_2$}}, \hat{i}, -\hat{P}_{\mbox{\tiny $i(k_1 k_2)$}})+\\
  &\hspace{.5cm}+M_3(\hat{i},k_1,-P_{\hat{i} k_1})\frac{1}{P_{\hat{i}k_1}^2}
    M_{3}(P_{\hat{i}k_1},k_2,-P_{\hat{i}(k_1 k_2)})+\\
  &\hspace{.5cm}+M_3(\hat{i},k_2,-P_{\hat{i}k_2})\frac{1}{P_{\hat{i}k_2}^2}
    M_{3}(P_{\hat{i}k_2},k_1,-P_{\hat{i}(k_1 k_2)}),
 \end{split}
\end{equation}
where the brackets in the notation $P_{i(k_1 k_2)}$ 
indicate that the momentum $p^{\mbox{\tiny $(k_1)$}}+p^{\mbox{\tiny $(k_2)$}}$ 
is on-shell in this limit, and the following identifications are understood: 
$\tilde{\lambda}^{(k_1 k_2)}\,=\,\tilde{\lambda}^{(k_1)}$ and
$\tilde{\lambda}^{(k_2)}\,=\,\frac{[k_2,\mu]}{[k_1,\mu]}\tilde{\lambda}^{(k_1)}$. In \eqref{lim3b} we wrote all the possible 
channels. 
Notice that for a $\delta$-derivative interaction $(\delta\,>\,0)$, one of the two three-point amplitudes in each of 
the terms on the r.h.s side of \eqref{lim3b} has to be anti-holomorphic. However, all the momenta in these three-particle 
amplitudes are characterised by having the anti-holomorphic spinors proportional to each other. As a consequence, such terms
are always zero.

In the case of $0$-derivative interactions, instead, both the holomorphic and anti-holomorphic term of a three-particle 
amplitude are relevant and all the terms in \eqref{lim3b} turn out to be non-zero.


As far as the factor $f_{i\mathcal{I}_k}^{(\nu,n)}$ is concerned, it is easy to see that in these cases the same
condition \eqref{z0K} holds, while the factors $f_{i k_1}^{(\nu,n)}$ and $f_{i k_2}^{(\nu,n)}$ are such that
\begin{equation}\eqlabel{z0K2}
 \lim_{[k_1,k_2]\rightarrow0}f_{i \bar{k}}^{(\nu,n)}\:\equiv\:
   \lim_{[k_1,\,k_2]\,\rightarrow\,0}\prod_{l=1}^{\nu+1}
    \left(1-\frac{P_{i\bar{k}}^2}{P_{i\bar{k}}^2(z_0^{(l)})}\right)\:=\:
   \prod_{l=1}^{\nu+1}\left(1-\frac{P_{i(k_1 k_2)}^2}{P_{i(k_1 k_2)}^2(\bar{z}_0^{(l)})}\right)\:\equiv\:
   f_{i(k_1 k_2)}^{(\nu,n-1)},
\end{equation}
where $\bar{z}_0$ indicates the zeroes of the lower point amplitudes and $\bar{k}\,\in\,\{k_1,\,k_2\}$.

The same reasoning applies in the case $k_1,\,k_2\,\in\,\mathcal{J}_k$ and $\tilde{\mathcal{J}}_k\,=\,\emptyset$ as 
$\langle k_1,\,k_2\rangle\rightarrow0$.

\subsubsection{Particles $k_1$ and $k_2$ on different sub-amplitudes}\label{subsub:diff}

Another difference with respect to the case analysed in section \ref{subsec:PK} lies on the fact that, in the present 
case, the class of terms 
$\left(\mathcal{K}_{\mbox{\tiny $\mathcal{I}$}},\mathcal{K}_{\mbox{\tiny $\mathcal{J}$}}\right)\,\neq\,
 \left(\emptyset,\emptyset\right)$ does contribute. Specifically, such a class of terms contains all 
those diagrams such that 
$\left(\mathcal{K}_{\mbox{\tiny $\mathcal{I}$}},\mathcal{K}_{\mbox{\tiny $\mathcal{J}$}}\right)\,=\,
 \left(k_1,k_2\right)$ or
$\left(\mathcal{K}_{\mbox{\tiny $\mathcal{I}$}},\mathcal{K}_{\mbox{\tiny $\mathcal{J}$}}\right)\,=\,
 \left(k_2,k_1\right)$. 
Obviously, the treatment of these two sub-cases is equivalent and related by relabelling $k_1\,\leftrightarrow\,k_2$.
Here we can further distinguish three sub-classes of diagrams: 
\begin{equation}\eqlabel{lim3a}
 \left(\tilde{\mathcal{I}}_k,\tilde{\mathcal{J}}_k\right)\,=\,
 \left\{\left(\tilde{\mathcal{I}}_k,\emptyset\right),\,\left(\emptyset,\tilde{\mathcal{J}}_k\right),
 \left(\tilde{\mathcal{I}}_k,\tilde{\mathcal{J}}_k\right)\right\}.
\end{equation}
The last case is analogous to the
$\left(\mathcal{K}_{\mbox{\tiny $\mathcal{I}$}}, \mathcal{K}_{\mbox{\tiny $\mathcal{J}$}}\right)$ one in
section \ref{subsec:PK}: for these diagrams $P_{\mbox{\tiny $k_1 k_2$}}^2\,\rightarrow\,0$ does not appear as a 
singularity. 

As far as the other two sub-classes of terms are concerned, they are characterised by the presence of three-particle 
amplitudes:
$M_{\mbox{\tiny R}}^{\mbox{\tiny $(i,j)$}}(\hat{P}_{\mbox{\tiny $i\mathcal{I}_k$}},\bar{k},\hat{j})$ for 
$\left(\tilde{\mathcal{I}}_k,\tilde{\mathcal{J}}_k\right)\,=\,\left(\tilde{\mathcal{I}}_k,\emptyset\right)$, and
$M_{\mbox{\tiny L}}^{\mbox{\tiny $(i,j)$}}(\hat{i},\bar{k},-\hat{P}_{\mbox{\tiny $i\mathcal{I}_k$}})$ for
$\left(\tilde{\mathcal{I}}_k,\tilde{\mathcal{J}}_k\right)\,=\,\left(\emptyset,\tilde{\mathcal{J}}_k\right)$,
where $\bar{k}\,\in\,\left\{k_1,\,k_2\right\}$. This implies that, for each of these two classes, there are
two contributions and they are such that $\bar{k}\,\cdot\,\hat{P}_{\mbox{\tiny $i\mathcal{I}_k$}}\,=\,0$.

For the case 
$\left(\tilde{\mathcal{I}}_k,\tilde{\mathcal{J}}_k\right)\,=\,\left(\tilde{\mathcal{I}}_k,\emptyset\right)$,
the BCFW-deformation on the spinors is provided in \eqref{BCFWdef2}, and the spinors evaluated at the location of the 
pole $z_{j \bar{k}}\,=\,-\langle j,\bar{k}\rangle/\langle i,\bar{k}\rangle$ are given by
\begin{equation}\eqlabel{lim3c}
 \begin{split}
 &\hat{\tilde{\lambda}}^{\mbox{\tiny $(i)$}}\:=\:\tilde{\lambda}^{\mbox{\tiny $(i)$}}+
   \frac{\langle j,\bar{k}\rangle}{\langle i,\bar{k}\rangle}\tilde{\lambda}^{\mbox{\tiny $(j)$}},
  \qquad
  \hat{\lambda}^{\mbox{\tiny $(j)$}}\:=\:\lambda^{\mbox{\tiny $(j)$}}-
   \frac{\langle j,\bar{k}\rangle}{\langle i,\bar{k}\rangle}\lambda^{\mbox{\tiny $(i)$}}\,=\,
   \frac{\langle i,j\rangle}{\langle i,\bar{k}\rangle}\lambda^{\mbox{\tiny $(\bar{k})$}},\\
 &-\hat{P}_{\mbox{\tiny $i\mathcal{I}_k$}}\:\equiv\:\hat{P}_{j\bar{k}}\:=\:P_{j\bar{k}}-
   \frac{\langle j,\bar{k}\rangle}{\langle i,\bar{k}\rangle}\lambda^{\mbox{\tiny $(i)$}}
    \tilde{\lambda}^{\mbox{\tiny $(j)$}}\:=\:
   \lambda^{\mbox{\tiny $(\bar{k})$}}\left(\tilde{\lambda}^{\mbox{\tiny $(\bar{k})$}}+
   \frac{\langle i,j\rangle}{\langle i,\bar{k}\rangle}\tilde{\lambda}^{\mbox{\tiny $(j)$}}\right).
 \end{split}
\end{equation}
For theories with $\delta$-derivative three-particle interaction ($\delta\,>\,0$), the proportionality among
the holomorphic spinors in $M_{\mbox{\tiny R}}^{\mbox{\tiny $(i,j)$}}(\hat{P}_{\mbox{\tiny $i\mathcal{I}_k$}},k_2,\hat{j})$
implies that it is expressed just in terms of the anti-holomorphic spinors \cite{Benincasa:2007xk}. For theories
with $0$-derivative three-particle interaction ($\delta\,=\,0$), both the holomorphic and anti-holomorphic term
in the three-particle amplitudes are non-zero \cite{Benincasa:2007xk} and the effective three-particle coupling
constant is just a linear combination of the holomorphic and anti-holomorphic ones, as showed earlier.

The limit $\langle k_1,\,k_2\rangle\:\longrightarrow\:0$ induces a singularity in the sub-amplitude 
$M_{\mbox{\tiny L}}^{\mbox{\tiny $(i,j)$}}(\hat{i},\tilde{k},\tilde{\mathcal{I}}_k,-\hat{P}_{i\mathcal{I}_k})$:
\begin{equation}\eqlabel{sing1}
 \begin{split}
  &\frac{1}{\left(p^{\mbox{\tiny $(k_1)$}}+\hat{P}_{j k_2}\right)^2}\:=\:
   \frac{1}{\langle k_1,\,k_2\rangle\left([k_1, k_2]+\frac{\langle i,j\rangle}{\langle i,k_2\rangle}[k_1,j]\right)},\\
  &\frac{1}{\left(p^{\mbox{\tiny $(k_2)$}}+\hat{P}_{j k_1}\right)^2}\:=\:
   \frac{1}{\langle k_2,\,k_1\rangle\left([k_2, k_1]+\frac{\langle i,j\rangle}{\langle i,k_1\rangle}[k_2,j]\right)}.
 \end{split}
\end{equation}
Notice that in the limit $\langle k_1,\,k_2\rangle\:\longrightarrow\:0$, the holomorphic spinors for $p^{(\tilde{k})}$
and $\hat{P}_{j\bar{k}}$ are proportional to each other, with $\tilde{k}\,\in\,\{k_1,\,k_2\}$ and 
$\tilde{k}\,\neq\,\bar{k}$.
This channel does not instead show a singularity in the limit $[k_1,\,k_2]\:\longrightarrow\:0$.
Repeating the same argument on the term characterised by 
$\left(\tilde{\mathcal{I}}_k,\tilde{\mathcal{J}}_k\right)\,=\,\left(\emptyset,\tilde{\mathcal{J}}_k\right)$, 
it turns out to be singular as $[k_1,\,k_2]\:\longrightarrow\:0$ rather than in the limit 
$\langle k_1,\,k_2\rangle\:\longrightarrow\:0$.

Therefore, in the limit $P_{\mbox{\tiny $k_1 k_2$}}^2\,\rightarrow\,0$, the sub-amplitudes 
$M_{\mbox{\tiny L}}^{\mbox{\tiny $(i,j)$}}(\hat{i},\tilde{k},\tilde{\mathcal{I}}_k,-\hat{P}_{i\mathcal{I}_k})$ and
\newline
$M_{\mbox{\tiny R}}^{\mbox{\tiny $(i,j)$}}(\hat{P}_{i\mathcal{I}_k}, \tilde{k}, \tilde{\mathcal{J}}_k, \hat{j})$
factorise as follows
\begin{equation}\eqlabel{lim3A}
 \begin{split}
  &M_{\mbox{\tiny L}}^{\mbox{\tiny $(i,j)$}}(\hat{i},\tilde{k},\tilde{\mathcal{I}}_k,\hat{P}_{j\bar{k}})\:
   \overset{\langle k_1, k_2\rangle\rightarrow0}{\longrightarrow}\:
   M_3(\tilde{k},\hat{P}_{\mbox{\tiny $j\bar{k}$}}, -P_{\mbox{\tiny $\tilde{k}(\hat{j}\bar{k})$}})
   \frac{1}{P_{\mbox{\tiny $\tilde{k}(\hat{j}\bar{k}$)}}^2}
   M_{\mbox{\tiny L}}^{\mbox{\tiny $(i,j)$}}(P_{\mbox{\tiny $\tilde{k}(\hat{j}\bar{k})$}},\hat{i},
 \tilde{\mathcal{I}}_k)
   \\
  &M_{\mbox{\tiny R}}^{\mbox{\tiny $(i,j)$}}(\hat{P}_{i\bar{k}}, \tilde{k}, \tilde{\mathcal{J}}_k, \hat{j})\:
   \overset{[k_1, k_2]\rightarrow0}{\longrightarrow}\:
   M_3(\tilde{k},\hat{P}_{\mbox{\tiny $i\bar{k}$}}, -P_{\mbox{\tiny $\tilde{k}(\hat{i}\bar{k})$}})
   \frac{1}{P_{\mbox{\tiny $\tilde{k}\hat{P}_{i\bar{k}}$}}^2}
  M_{\mbox{\tiny R}}^{\mbox{\tiny $(i,j)$}}(P_{\mbox{\tiny $\tilde{k}(\hat{i}\bar{k})$}}, \tilde{\mathcal{J}}_k, \hat{j}).
 \end{split}
\end{equation}

As far as the factor $f_{i\mathcal{I}_k}^{(\nu,n)}$ is concerned, it is easy to see that in the above limits one has
\begin{equation}\eqlabel{lim3B}
 \begin{split}
  &\lim_{\langle k_1,\,k_2\rangle\,\rightarrow\,0}f_{j\bar{k}}^{(\nu,n)}\:\equiv\:
   \lim_{\langle k_1,\,k_2\rangle\,\rightarrow\,0}\prod_{l=1}^{\nu+1}
    \left(1-\frac{P_{j\bar{k}}^2}{P_{j\bar{k}}^2(z_0^{(l)})}\right)\:=\:
   \prod_{l=1}^{\nu+1}\left(1-\frac{P_{j(k_1 k_2)}^2}{P_{j(k_1 k_2)}^2(\bar{z}_0^{(l)})}\right)\:\equiv\:
   f_{j(k_1 k_2)}^{(\nu,n-1)},\\
  &\lim_{[k_1,\,k_2]\,\rightarrow\,0}f_{i\bar{k}}^{(\nu,n)}\:\equiv\:
   \lim_{[k_1,\,k_2]\,\rightarrow\,0}\prod_{l=1}^{\nu+1}
    \left(1-\frac{P_{i\bar{k}}^2}{P_{i\bar{k}}^2(z_0^{(l)})}\right)\:=\:
   \prod_{l=1}^{\nu+1}\left(1-\frac{P_{i(k_1 k_2)}^2}{P_{i(k_1 k_2)}^2(\bar{z}_0^{(l)})}\right)\:\equiv\:
   f_{i(k_1 k_2)}^{(\nu,n-1)},
 \end{split}
\end{equation}
where $\bar{z}_{0}^{(l)}$ indicates a zero of an $(n-1)$-particle amplitude. We will comment in more detail on this
in the next subsection.

\subsubsection{Factorisation in the collinear limit $P_{k_1 k_2}^2\,\rightarrow\,0$}\label{subsub:Fact}

In the previous subsections, we discussed the behaviour of the different classes of terms in the generalised on-shell 
representation
as the limit $P_{k_1 k_2}^2\,\rightarrow\,0$ is taken. Using this knowledge let us now see explicitly how the correct
factorisation property is reproduced.

First of all, let us notice that the $(n-1)$-particle amplitude in \eqref{Pk1k2lim} can be expressed via the on-shell
representation \eqref{Mfin}. For the case of theories with $0$-derivative interactions, it is easy to see that all
terms in \eqref{Pk1k2lim} are correctly reproduced by the classes of terms for $M_n$ discussed in the subsection
\ref{subsub:same}. More precisely, such classes contain extra-terms, which are of the type of the ones in second and third
lines of \eqref{lim3b}. This means that, together with the terms discussed in subsection \ref{subsub:diff}, they need to 
add up to zero, which is straightforward to check, with a little algebra.

\begin{figure}[htbp]
 \centering 
 \scalebox{.35}{\includegraphics{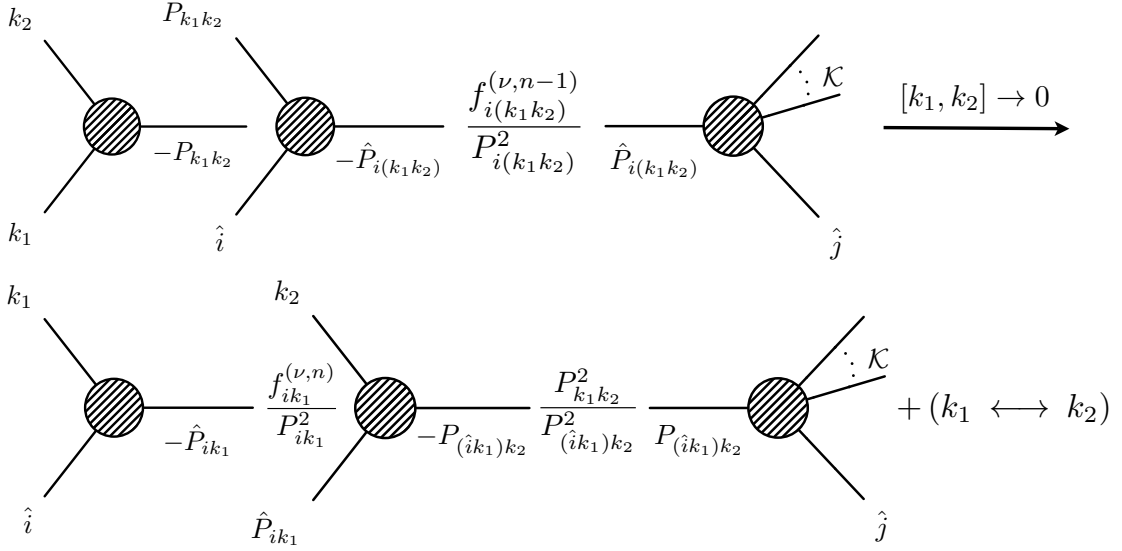}}
 \caption{Collinear limit $P_{k_1 k_2}^2\,\rightarrow\,0$. For theories with $\delta$-derivative interactions 
          ($\delta\,>\,0$), the analysis of this collinear limit returns a non-trivial equality which, for theories
          with internal quantum numbers, gives the algebra of the internal group \eqref{Pk1k2lim3} and \eqref{Pk1k2lim4}}.
 \label{fig:Pk1k2}
\end{figure}

For theories with $\delta$-derivative interactions $(\delta\,>\,0)$, the on-shell representation \eqref{Mfin} of the
$n$-point amplitude gets a vanishing contribution from terms where the two particles $k_1$ and $k_2$ belong to the
same sub-amplitude and such sub-amplitude is a four-particle amplitude. However, expressing the $(n-1)$-amplitude in
\eqref{Pk1k2lim} through the recursion relation \eqref{Mfin}, the r.h.s. of \eqref{Pk1k2lim} actually shows such type of 
contributions. This means that they should be reproduced by the terms discussed in subsection \ref{subsub:diff}.
In the limit $[k_1,k_2]\,\rightarrow\,0$:
\begin{equation}\eqlabel{Pk1k2lim3}
 \begin{split}
  &M_{3}^{\mbox{\tiny H}}\left(k_1,k_2,-P_{k_1 k_2}\right)
  M_3^{\mbox{\tiny H}}\left(\hat{i},P_{k_1 k_2},-\hat{P}_{i(k_1 k_2)}\right)
   \frac{f_{i(k_1 k_2)}^{\mbox{\tiny $(\nu,n-1)$}}}{P_{i(k_1 k_2)}^2}
   M_{n-2}\left(\hat{P}_{i(k_1 k_2)},\mathcal{K},\hat{j}\right)\:=\\
  &=\:\lim_{[k_1,k_2]\rightarrow0}
   \left[
    M_{3}^{\mbox{\tiny H}}\left(\hat{i},k_1,-\hat{P}_{i k_1}\right)\frac{f_{i k_1}^{\mbox{\tiny $(\nu,n)$}}}{P_{i k_1}^2}
    M_{3}^{\mbox{\tiny H}}\left(\hat{P}_{i k_1},k_2,-P_{(\hat{i} k_1)k_2}\right)\frac{\langle k_1,k_2\rangle}{
     \langle k_1,k_2\rangle+\frac{[j,i]}{[j,k_1]}\langle i,k_2\rangle}\times
   \right.\\
  &\hspace{1cm}\times
   \left.
    M_{n-2}\left(P_{(\hat{i}k_1)k_2},\mathcal{K},\hat{j}\right)+\left(k_1\:\longleftrightarrow\:k_2\right)
   \right],
 \end{split}
\end{equation}
where we used the notation $M_3^{\mbox{\tiny H}}$ to emphasise that these three-particle amplitudes are functions
of the holomorphic spinors only (figure \ref{fig:Pk1k2}). 

Notice that, in such a limit, $P_{(\hat{i}k_1)k_2}\,\equiv\,\hat{P}_{i(k_1 k_2)}$ and, as discussed earlier,
$f_{i\bar{k}}^{\mbox{\tiny $(\nu,n)$}}\,\rightarrow\,f_{i(k_1 k_2)}^{\mbox{\tiny $(\nu,n-1)$}}$. As a consequence,
one can factor out $f_{i k_1 k_2}^{\mbox{\tiny $(\nu,n)$}}M_{n-2}\left(\hat{P}_{i(k_1 k_2)},\mathcal{K},\hat{j}\right)$ in
\eqref{Pk1k2lim3}. With a little of algebra, such a relation can be written as
\begin{equation}\eqlabel{Pk1k2lim4}
 \begin{split}
  \varepsilon_{\mbox{\tiny $k_1 k_2 a$}}\varepsilon_{\mbox{\tiny $i a P$}}(-1)^{2h_i}
   &\left(1+\frac{[k_2,\mu]}{[k_1,\mu]}\frac{\langle i,k_2\rangle}{\langle k_1,k_2\rangle}\right)^{\delta-1}=
  \varepsilon_{\mbox{\tiny $i k_1 a$}}f_{\mbox{\tiny $a k_2 P$}}(-1)^{2h_2}
   \left(1+\frac{[j,i]}{[j,k_1]}\frac{\langle i,k_2\rangle}{\langle k_1,k_2\rangle}\right)^{\delta-1}+\\
  &+\varepsilon_{\mbox{\tiny $i k_2 a$}}\varepsilon_{\mbox{\tiny $a k_1 P$}}(-1)^{2h_1+1}
   \left(\frac{[j,i]}{[j,k_1]}\frac{\langle i,k_2\rangle}{\langle k_1,k_2\rangle}+\frac{[k_2,\mu]}{[k_1,\mu]}
    \frac{\langle i,k_2\rangle}{\langle k_1,i\rangle}\right)^{\delta-1},
 \end{split}
\end{equation}
where we allowed for the possibility of internal quantum numbers by introducing the structure constants $\varepsilon$'s,
$\kappa\,\rightarrow\,\kappa\varepsilon_{abc}$, and a sum for the repeated indices is understood. In case there is no 
internal symmetry, one can set them to $1$. One can check in
both cases how this relation is satisfied by fixing the theory, {\it i.e.} the helicities of the particles involved 
$\{h_i,h_1,h_2\}$ and the number of the derivative in the three-particle interaction. 

The analysis of the limit $\langle k_1,k_2\rangle\rightarrow0$ proceeds along the same lines.

\subsection{The collinear limit $P_{ij}^2\,\rightarrow\,0$}\label{subsec:Pij}

A great deal of information is provided by the $\left(i,j\right)$-channel, which does not appear explicitly in \eqref{Mfin}.
As discussed in \cite{Schuster:2008nh} for scattering of gluons and gravitons, in the standard BCFW-representation this
singularity appears as a soft singularity, when the deformed momenta of either particle-$i$  or particle-$j$ vanishes.
We will show how requiring the correct factorisation in this channel fixes the complex-UV behaviour as well as it provides 
conditions on the zeroes. As for the $(k_1,\,k_2)$-channel, we analyse the limits $[i,\,j]\,\rightarrow\,0$ and 
$\langle i,\,j\rangle\,\rightarrow\,0$ separately.

Let us start with $[i,\,j]\,\rightarrow\,0$.
In this limit, the amplitude should factorise as follows
\begin{equation}\eqlabel{Pijlim1}
 \lim_{\mbox{\tiny $[i,j]$}\:\rightarrow\:0}
    P_{\mbox{\tiny $ij$}}^2\,M_n\:=\:
   M_{3}(i,\,j,\,-P_{ij}^{h_{ij}})\, M_{n-1}(P_{ij}^{-h_{ij}}, \mathcal{K}),
\end{equation}
with $h_{ij}\,=\,-h_{i}-h_{j}-\delta$. For future reference, it is convenient to write down here the explicit expression 
for the three-particle amplitude in \eqref{Pijlim1}:
\begin{equation}\eqlabel{M3ij}
 M_{3}(i,\,j,\,-P_{ij}^{h_{ij}})\:=\:\kappa_{1-\delta}\left(-1\right)^{2(h_i+h_j)}\langle i,j\rangle^{\delta}
 \left(\frac{[i,\mu]}{[j,\mu]}\right)^{2h_i+\delta},
\end{equation}
where we used the fact that, in this limit, the on-shell momentum $P_{ij}$ can be written as
\begin{equation}\eqlabel{PijOnShell}
P_{ij}\,=\,\left(\frac{[i,\mu]}{[j,\mu]}\lambda^{(i)}+\lambda^{(j)}\right)\tilde{\lambda}^{(j)},
\end{equation}
with the identifications
\begin{equation}\eqlabel{PijId}
 \tilde{\lambda}^{(i)}\:\overset{\mbox{\tiny def}}{=}\:\frac{[i,\mu]}{[j,\mu]}\tilde{\lambda}^{(j)},\quad
 \lambda^{(ij)}\:\overset{\mbox{\tiny def}}{=}\:\frac{[i,\mu]}{[j,\mu]}\lambda^{(i)}+\lambda^{(j)}, \quad
 \tilde{\lambda}^{(ij)}\:\overset{\mbox{\tiny def}}{=}\:\tilde{\lambda}^{(j)},
\end{equation}
and $\tilde{\lambda}^{(\mu)}$ being some reference spinor. The expression for the three-particle amplitude \eqref{M3ij}
is valid for any $\delta\,\ge\,0$.

\begin{figure}[htbp]
 \centering
 \scalebox{.30}{\includegraphics{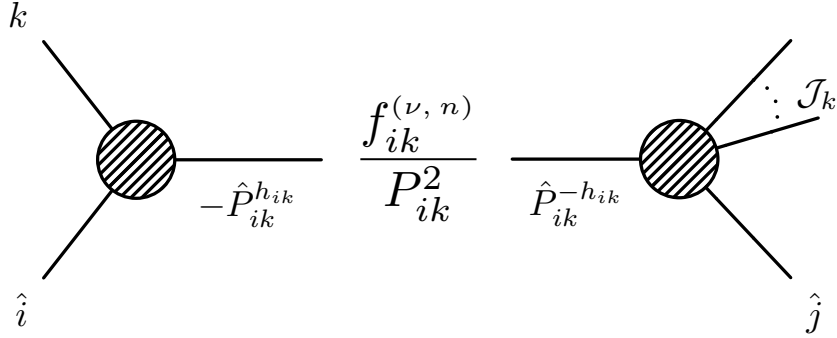}}
 \caption{Collinear limit $[i,j]\,\rightarrow\,0$. There are just one class of terms contributing to this limit,
          which is characterised by a three-particle amplitude of type $M_{3}\left(\hat{i},k,-\hat{P}_{ij}\right)$.}
 \label{fig:Pij1}
\end{figure}

It is easy to see that the only terms of the recursion relation which
can contribute are (see figure \ref{fig:Pij1})
\begin{equation}\eqlabel{limij1}
 \lim_{\mbox{\tiny $[i,\,j]\,\rightarrow\,0$}}P_{ij}^2\,M_n\:=\:
 \lim_{\mbox{\tiny $[i,\,j]\,\rightarrow\,0$}}P_{ij}^2\sum_{k}M_3(\hat{i},\,k,\,-\hat{P}_{ik}^{h_{ik}})
  \frac{f_{ik}^{\mbox{\tiny $(\nu,n)$}}}{P_{ik}^2}M_{n-1}(\hat{P}_{ik}^{-h_{ik}},\,\mathcal{J}_k,\,\hat{j}),
\end{equation}
where $h_{ik}\,=\,-h_{i}-h_{k}-\delta$, the poles are $z_{ik}\,=\,[i,\,k]/[j,\,k]$ and the relevant quantities computed at
the location of the poles are
\begin{equation}\eqlabel{limij2}
 \hat{\tilde{\lambda}}^{\mbox{\tiny (i)}}\:=\:\frac{[j,i]}{[j,k]}\tilde{\lambda}^{\mbox{\tiny (k)}},\quad
 \hat{\lambda}^{\mbox{\tiny (j)}}\:=\:\lambda^{\mbox{\tiny (j)}}+\frac{[i,k]}{[j,k]}\lambda^{\mbox{\tiny (i)}},\quad
 \hat{P}_{ik}\:=\:p^{\mbox{\tiny $(k)$}}+\frac{[j,i]}{[j,k]}\lambda^{\mbox{\tiny (i)}}\tilde{\lambda}^{\mbox{\tiny (k)}}.
\end{equation}

From \eqref{limij2}, one can notice that the limit $[i,j]\,\rightarrow\,0$ implies that
$\hat{\tilde{\lambda}}^{\mbox{\tiny $(i)$}}$, and consequently the momentum of particle-$i$, vanishes, as well as
$\hat{p}^{\mbox{\tiny $(j)$}}\,\rightarrow\,P_{ij}$ and $\hat{P}_{ik}\,\rightarrow\,p^{\mbox{\tiny $(k)$}}$. A
further consequence is that all the $(n-1)$-particle amplitudes in the sum \eqref{limij1} are mapped into
$M_{n-1}\left(P_{ij}^{h_{j}},p_{k}^{-h_{ik}},\mathcal{J}_{k}\right)$. There is an important subtlety. This
$(n-1)$-particle amplitude has exactly the same momenta for the external states as the one in \eqref{Pijlim1}, but
with the fundamental difference that, generically, the helicities carried by the states with momenta $P_{ij}$ and
$p_k$ are not the same as the states with same momenta in \eqref{Pijlim1}. However, it is always possible to relate
these amplitudes by a dimensionless factor:
\begin{equation}\eqlabel{helfactij}
 M_{n-1}\left(P_{ij}^{h_{j}},p_{k}^{-h_{ik}},\mathcal{J}_{k}\right)\:=\:
  \mathcal{H}_{n-1}^{(k)}M_{n-1}(P_{ij}^{-h_{ij}}, \mathcal{K}),
\end{equation}
with $\mathcal{H}_{n-1}^{(k)}\,=\,1$ if $h_{ij}\,=\,-h_j$ and $h_{ik}\,=\,-h_{k}$. Beyond being dimensionless,
this factor can depend in an helicity-blind way on the spinors of the particles in $\mathcal{J}_k$ and it has to
show the proper helicity scaling with the spinors of $P_{ij}$ and $p_{k}$. Using the relation \eqref{helfactij},
the $(n-1)$-particle amplitude can be factored out from the sum \eqref{limij1}, leaving in the sum the factor 
$\mathcal{H}_{n-1}^{(k)}$, which can be computed explicitly for a given theory.
Let us now focus on the three-particle amplitudes $M_3(\hat{i},\,k,\,-\hat{P}_{ik}^{h_{ik}})$, which can be explicitly
written as
\begin{equation}\eqlabel{M3ik}
 M_3(\hat{i},\,k,\,-\hat{P}_{ik}^{h_{ik}})\:=\:\mathcal{\kappa}_{1-\delta}(-1)^{2(h_i+h_k)}
  \langle i,k\rangle^{\delta}\left(\frac{[j,i]}{[j,k]}\right)^{2h_i+\delta}.
\end{equation}
From \eqref{M3ij}, \eqref{helfactij} and \eqref{M3ik}, eq \eqref{limij1} becomes
\begin{equation}\eqlabel{limij6}
 \begin{split}
 \lim_{\mbox{\tiny $[i,\,j]\,\rightarrow\,0$}}P_{ij}^2\,M_n\:=\:&
  \left[\lim_{[i,j]\rightarrow0}\sum_{k}(-1)^{2(h_j+h_k)}
  \left(\frac{\langle i,k\rangle}{\langle i,j\rangle}\right)^{\delta-1}
  \left(\frac{[j,i]}{[j,k]}\frac{[j,\mu]}{[i,\mu]}\right)^{2h_i+\delta}\frac{[i,j]}{[i,k]}f_{ik}^{(\nu,n)}
  \mathcal{H}_{n-1}^{(k)}\right]\times\\
 &\times M_3\left(i,j,-P_{ij}\right)M_{n-1}(P_{ij},\mathcal{K}).
 \end{split}
\end{equation}
The above expression reproduces the correct factorisation property \eqref{Pijlim1} if and only if the term in square 
brackets is one. Let us now analyse it in some detail. First of all, one notices the presence of the factor
$[i,j]^{2h_i+\delta+1}$. 

If $2h_i+\delta+1\,>\,0$\footnote{Notice that the inequality is due to the fact that
the sum in \eqref{limij6} might generate an extra factor of $[i,j]$ at the numerator, as happens for graviton amplitudes
under the standard BCFW deformation \cite{Schuster:2008nh}.}, the requirement of $f_{ik}^{(\nu,n)}$ to be proportional to 
some negative power of $[i,j]$ needs necessarily to hold in order to reproduce the correct factorisation properties.

Let us now look at the explicit expression for $f_{ik}^{\mbox{\tiny $(\nu,n)$}}$ when $\nu\,\ge\,0$. Unitarity, through the 
requirement for the amplitude to factorise properly, implies that
\begin{equation}\eqlabel{z0cond}
 P_{ik}^{2}(z_{0}^{\mbox{\tiny $(l)$}})\:\equiv\:\langle i,k\rangle\left([i,k]-z_{0}^{\mbox{\tiny $(l)$}}[j,k]\right)\:=\:
  \langle i,k\rangle\alpha_{ik}^{\mbox{\tiny $(l)$}}[i,j],
\end{equation}
and, consequently, $f_{ik}^{\mbox{\tiny $(\nu,n)$}}$ with $\nu\,\ge\,0$ becomes
\begin{equation}\eqlabel{fik}
 \begin{split}
 f_{ik}^{\mbox{\tiny $(\nu,n)$}}\:&\equiv\:\prod_{l=1}^{\nu+1}
  \left(1-\frac{P_{ik}^2}{P_{ik}^2 (z_{0}^{\mbox{\tiny $(l)$}})}\right)
  \:\equiv\:
  (-1)^{\nu+1}\left(\prod_{l=1}^{\nu+1}\frac{P_{ik}^2}{P_{ik}^2 (z_{0}^{\mbox{\tiny $(l)$}})}\right)
  \prod_{l=1}^{\nu+1}\left(1-\frac{P_{ik}^2 (z_{0}^{\mbox{\tiny $(l)$}})}{P_{ik}^2}\right)\:=\\
  &=\:
  (-1)^{\nu+1}\left(\frac{[i,k]}{[i,j]}\right)^{\nu+1}\left(\prod_{l=1}^{\nu+1}\alpha_{ik}^{\mbox{\tiny $(l)$}}\right)^{-1}
   \prod_{l=1}^{\nu+1}\left(1-\alpha_{ik}^{\mbox{\tiny $(l)$}}\frac{[i,j]}{[i,k]}\right).
 \end{split}
\end{equation}
From \eqref{fik}, the collinear limit \eqref{limij6} can be conveniently written as
\begin{equation}\eqlabel{limij3}
 \begin{split}
  \lim_{\mbox{\tiny $[i,\,j]\,\rightarrow\,0$}}P_{ij}^2\,M_n\:=\:&
  \left[\lim_{[i,j]\rightarrow0}\sum_{k}(-1)^{\xi}
  \left(\frac{\langle i,k\rangle}{\langle i,j\rangle}\right)^{\delta-1}
  \left(\frac{[i,k]}{[j,k]}\frac{[j,\mu]}{[i,\mu]}\right)^{2h_i+\delta}
  \left(\frac{[i,j]}{[i,k]}\right)^{2h_i+\delta-\nu}
  \frac{\mathcal{H}_{n-1}^{(k)}}{\prod_{l=1}^{\nu+1}\alpha_{ik}^{\mbox{\tiny $(l)$}}}\right]\times\\
 &\times M_3\left(i,j,-P_{ij}\right)M_{n-1}(P_{ij},\mathcal{K}),
 \end{split}
\end{equation}
where $\xi\,=\,2(h_i+h_j+h_k)+\delta+\nu+1$. Notice that, in this limit, the term in the second round-brackets in
\eqref{limij3} is actually one. Thus, the correct factorisation requirement in the $(i,j)$-channel can be finally written
as
\begin{equation}\eqlabel{limij4}
  \lim_{[i,j]\rightarrow0}\sum_{k}(-1)^{2(h_i+h_j+h_k)+\delta+\nu+1}
   \left[
  \left(\frac{\langle i,k\rangle}{\langle i,j\rangle}\right)^{\delta-1}
 \left(\frac{[i,j]}{[i,k]}\right)^{2h_i+\delta-\nu}
  \frac{\mathcal{H}_{n-1}^{(k)}}{\prod_{l=1}^{\nu+1}\alpha_{ik}^{\mbox{\tiny $(l)$}}}\right]
 \:=\:1.
\end{equation}
The condition \eqref{limij4} univocally fixes $\nu$, and therefore the large-$z$ behaviour of the amplitudes, for
a given theory:
\begin{equation}\eqlabel{ComplexUV}
\nu\,=\,2h_i+\delta+m,
\end{equation}
where $m$ can be one or zero, depending on the fact that the sum \eqref{limij6} generates or not one extra factor of 
$[i,j]$. However, in order to check whether this might or might not be the case, one would need to look at specific
theories. It is important to point out that one can see that such an extra factor does not depend on the number $n$ of the 
external states.
Furthermore, the large-$z$ behaviour $\nu$ depends just on the helicities of the particles whose momenta have been deformed
as well as on the number of derivatives of the three-particle interactions.

If, instead, $2h_i+\delta+1\,\le\,0$, the term in square brackets in \eqref{limij6} is finite and 
different from zero for $f_{ik}^{(\nu,n)}\,=\,1$, {\it i.e.} the standard BCFW recursion relation holds. Thus the helicity 
of particle-$i$ - whose anti-holomorphic spinor has been deformed - needs to be 
negative in any case. Just to confirm this, let us discuss the behaviour of the amplitude as 
$[i,j]\,\rightarrow\,0$ and $h_i$ is positive. There are two cases to take into consideration which are related to the two 
possible $2$-particle deformations: 
$\left(h_{i},h_{j}\right)\,=\,\left\{\left(+,+\right),\left(+,-\right)\right\}$. 

For $\delta\,>\,0$, in the first case
no factorisation should occur and indeed this is the case given that the eventual three-particle amplitude in \eqref{Pijlim1}
would have to be anti-holomorphic and, therefore, vanishing in the limit $[i,j]\,\rightarrow\,0$. In the second
case, instead, the correct factorisation is not reproduced. This just implies that the standard BCFW recursion relation
does not hold and $\nu\,\ge\,0$. Notice, in fact, that the momenta deformation \eqref{BCFWdef2} with helicities 
$\left(h_{i},h_{j}\right)\,=\,\left(+,-\right)$ is what in the literature has been referred to as ``wrong shift''.
In such a case, the correct factorisation is reproduced if and only if the ``weight'' 
$f_{ik}^{\mbox{\tiny $(\nu,n)$}}$ is proportional to some negative power of $[i,j]$ in such a way to cancel the 
$[i,j]$-term in the numerator. 

For theories with $0$-derivative three-particle interactions, in both cases the factorisation is allowed, meaning that,
as before, $f_{ik}^{\mbox{\tiny $(\nu,n)$}}$ needs to be proportional to some negative power of $[i,j]$ and, therefore, 
$\nu\,\ge\,0$.

\begin{figure}[htbp]
 \centering
 \scalebox{.35}{\includegraphics{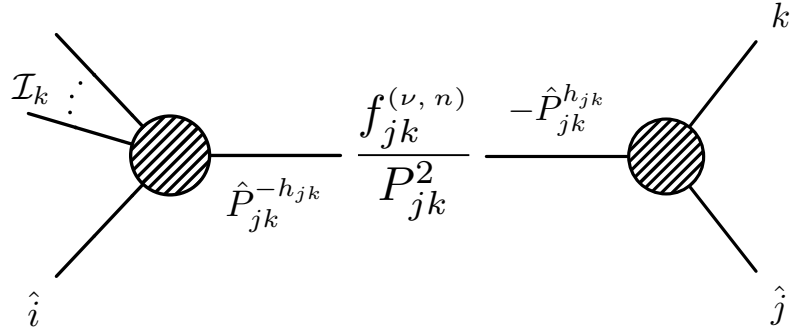}}
 \caption{Collinear limit $\langle i,j\rangle\,\rightarrow\,0$. There are just one class of terms contributing to this limit,
          which is characterised by a three-particle amplitude of type $M_{3}\left(\hat{P}_{ij},k,\hat{j}\right)$.}
 \label{fig:Pij2}
\end{figure}

Following the same arguments above, one can discuss the holomorphic limit $\langle i,j\rangle\,\rightarrow\,0$ 
(figure \ref{fig:Pij2}),
obtaining the conditions
\begin{equation}\eqlabel{limij5}
 \begin{split}
  &\hspace{4.5cm} P_{jk}^2(z_0^{\mbox{\tiny $(l)$}})\:=\:\langle i,j\rangle\alpha_{jk}^{\mbox{\tiny $(l)$}}[j,k],\\
  &\lim_{\langle i,j\rangle\rightarrow0}\sum_k(-1)^{2(h_i+h_j+h_k)+\delta+\nu+1}
    \left[
     \left(\frac{[k,j]}{[i,j]}\right)^{\delta-1}
     \left(\frac{\langle i,j\rangle}{\langle j,k\rangle}\right)^{\delta-2h_j-\nu}
     \frac{\tilde{\mathcal{H}}_{n-1}^{(k)}}{\prod_{l=1}^{\nu+1}\alpha_{jk}^{\mbox{\tiny $(l)$}}}
    \right]\:=\:1,
 \end{split}
\end{equation}
and the factor $\tilde{\mathcal{H}}_{n-1}^{(k)}$ is defined analogously to $\mathcal{H}_{n-1}^{(k)}$ in \eqref{helfactij}.

Notice that two conditions \eqref{limij4} and \eqref{limij5} do not need to hold simultaneously since a given theory may
factorise just under one of the two limits.

Summarising, the analysis of the $P_{ij}^2\,\rightarrow\,0$ limit shows that the the large-$z$ behaviour is independent of 
the number of external states and it depends only on the characteristics of the interactions and on the helicities of the
deformed particles.

\subsection{Collinear/Multi-particle limits and zeroes of the amplitudes: a Summary}\label{subsec:Sum}

In the previous subsections we have seen how unitarity, through the analysis of the collinear/multi-particle limits, fixes
conditions on the zeroes of the amplitudes or, more precisely, on the Lorentz invariants computed at the location of the
zeroes. For the sake of clarity, it is suitable to summarise here these conditions:
\begin{equation}\eqlabel{zeroes-sum}
 \begin{split}
  &\hspace{2.5cm}P_{ik}^2(z_0^{\mbox{\tiny $(l)$}})\:=\:\langle i,k\rangle\alpha_{ik}^{\mbox{\tiny $(l)$}}[i,j],\qquad
   P_{jk}^2(z_0^{\mbox{\tiny $(l)$}})\:=\:\langle i,j\rangle\alpha_{jk}^{\mbox{\tiny $(l)$}}[j,k],\\
  &\hspace{2.5cm}\lim_{P_{\mathcal{K}}^2\rightarrow0}f_{\mbox{\tiny $i\mathcal{I}_k$}}^{\mbox{\tiny $(\nu,\,n)$}}\:=\:
    f_{\mbox{\tiny $i\mathcal{I}_k$}}^{\mbox{\tiny $(\nu,\,n-s+1)$}},\qquad
   \lim_{P_{i\mathcal{I}_k}^2\rightarrow0}f_{i\mathcal{I}_k}^{\mbox{\tiny $(\nu,n)$}}\,=\,1,\\
  &\hspace{2.5cm}\lim_{[k_1,k_2]\rightarrow0}f_{\mbox{\tiny $i\bar{k}$}}^{\mbox{\tiny $(\nu,n)$}}\:=\:
    f_{\mbox{\tiny $i(k_1 k_2)$}}^{\mbox{\tiny $(\nu,n-1)$}},\qquad
   \lim_{\langle k_1,k_2\rangle\rightarrow0}f_{\mbox{\tiny $j\bar{k}$}}^{\mbox{\tiny $(\nu,n)$}}\:=\:
    f_{\mbox{\tiny $j(k_1 k_2)$}}^{\mbox{\tiny $(\nu,n-1)$}},\\
  &\lim_{[i,j]\rightarrow0}\sum_{k}(-1)^{2(h_i+h_j+h_k)+\delta+\nu+1}
   \left[
    \left(\frac{\langle i,k\rangle}{\langle i,j\rangle}\right)^{\delta-1}
    \left(\frac{[i,j]}{[i,k]}\right)^{2h_i+\delta-\nu}
    \frac{\mathcal{H}_{n-1}^{(k)}}{\prod_{l=1}^{\nu+1}\alpha_{ik}^{\mbox{\tiny $(l)$}}}\right]\:=\:1,\\
  &\lim_{\langle i,j\rangle\rightarrow0}\sum_k(-1)^{2(h_i+h_j+h_k)+\delta+\nu+1}
 \left[
  \left(\frac{[k,j]}{[i,j]}\right)^{\delta-1}
  \left(\frac{\langle i,j\rangle}{\langle j,k\rangle}\right)^{\delta-2h_j-\nu}
  \frac{\tilde{\mathcal{H}}_{n-1}^{(k)}}{\prod_{l=1}^{\nu+1}\alpha_{jk}^{\mbox{\tiny $(l)$}}}
 \right]\:=\:1.
 \end{split}
\end{equation}

Some comments are now in order. The collinear/multi-particle limits relate as well the ``weight'' factors
$f_{\mbox{\tiny $i\mathcal{I}_k$}}^{\mbox{\tiny $(\nu,n)$}}$ to the ones of amplitudes with fewer external states
through the relations \eqref{zeroes-sum}. This is a hint that it should be possible to reconstruct such factors from
the ones of smaller amplitudes. In some cases, such a connection driven by \eqref{zeroes-sum} is obvious and we 
will discuss them  in section \ref{sec:Ex}. However, one would like to formalise a consistent procedure to find it.
Even if some attempts, {\it e.g.} looking at a second BCFW-deformation, seemed to point towards the correct direction, 
so far we did not succeed and we leave it for future work. The existence of such a connection would imply that the last two
relations in \eqref{zeroes-sum} for the four-particle amplitudes would univocally fix the ``weights'' 
$f_{\mbox{\tiny $i\mathcal{I}_k$}}^{\mbox{\tiny $(\nu,n)$}}$ for any $n$. These relations are easy to solve in the case of
four particles, and lead to the following conditions on the zeroes \cite{Benincasa:2011bc}:
\begin{equation}\eqlabel{4ptzeroes}
 \prod_{r=1}^{N_{P}^{\mbox{\tiny fin}}}P_{r}^2(z_0^{\mbox{\tiny $(l)$}})\:=\:
  (-1)^{N_{P}^{\mbox{\tiny fin}}}\left(P_{ij}^2\right)^{N_{P}^{\mbox{\tiny fin}}},
\end{equation}
where $N_{P}^{\mbox{\tiny fin}}$ is the number of BCFW-poles, $P_r^2(z_0^{\mbox{\tiny $(l)$}})$ is the Mandelstam variable 
related to the pole $r$ and evaluated at the location of the zero $z_0^{\mbox{\tiny $(l)$}}$\footnote{For a more
extensive discussion about the condition \eqref{4ptzeroes} on the zeroes and, more generally, about the four-particle
amplitudes, see \cite{Benincasa:2011bc}}.

\section{More on the complex-UV limit}\label{sec:complexUV}

In section \ref{subsec:Pij} we have seen how the large-$z$ behaviour for the amplitudes can be determined by the analysis
of the collinear limit $P_{ij}^2\,\rightarrow\,0$ and is independent of the number of external states. A more extensive
discussion about this is in order.

Already in \cite{Schuster:2008nh}, it was noticed that the correct collinear singularity in the $(i,j)$-channel emerges
in the BCFW construction as a soft singularity $\hat{p}^{\mbox{\tiny $(i)$}}\,\rightarrow\,0$ or 
$\hat{p}^{\mbox{\tiny $(j)$}}\,\rightarrow\,0$ and, moreover, that the standard BCFW recursion relations were failing when
this soft singularity was not enough to reproduce the correct factorisation property in this channel. 

The analysis in \cite{Schuster:2008nh} as well as the one in this paper point out a connection between the soft limits for
the deformed particles / factorisation in the $(i,j)$-channel and the large-$z$ behaviour. In section \ref{subsec:Pij},
we showed how the large-$z$ behaviour is fixed just in terms of the number of derivatives of the particular interaction and
the helicities of the deformed particles. The same exact analysis, which is valid for arbitrary $n$, can be done for
amplitudes which satisfy standard BCFW recursion relations. Conditions of the type of those in the last two lines in 
\eqref{zeroes-sum} link the helicities of the deformed particles to the number of derivatives in the three-particle 
interactions. Specifically,
\begin{equation}\eqlabel{BCFWhihj}
 h_{i}\,=\,-\frac{\delta+1+m}{2} \quad\mbox{and/or}\quad h_{j}\,=\,\frac{\delta+1+m}{2},
\end{equation}
where $m$ can be one or zero, depending on the fact that the sum of the terms contributing in the 
$P_{ij}^2\,\rightarrow\,0$ generates or not one extra factor of $[i,j]$ or $\langle i,j\rangle$.

\section{Examples}\label{sec:Ex}

In this section we apply the generalised on-shell recursion relations \eqref{Mfin} and the conditions \eqref{zeroes-sum}
for constructing the scattering amplitudes for a number of examples.

\subsection{Gluon scattering amplitudes: ``wrong'' deformation}\label{subsec:Gluon}

It is well known that there are three classes of deformations \eqref{BCFWdef2} under which the standard BCFW recursion
relations are valid for gluon scattering amplitudes: 
$$(h_i,h_j)\,=\,\left\{(-,+),\,(-,-),\,(+,+)\right\}.$$
A further possibility for the choice of the helicities $(h_i,\,h_j)\,=\,(+,-)$ leads the deformed amplitude 
$M_{n}^{\mbox{\tiny $(i,j)$}}$ not to vanish as $z\,\rightarrow\,0$. Using the recursion relations \eqref{Mfin}, we 
reconstruct the gluons amplitudes through the ``wrong'' deformation. This issue was discussed in \cite{Feng:2010ku},
where the boundary term for gluon amplitudes has been obtained through on-shell recursion relations from 
$\mathcal{N}=4$ Super Yang-Mills.

\subsubsection{MHV amplitudes}\label{subsubsec:GluonMHV}

Let us start with the simplest example, the MHV gluon $n$-particle scattering amplitude
$M_{n}(1^{-},2^{+},3^{-},4^{+},\ldots,n^{+})$, using the deformation
\begin{equation}\eqlabel{BCFWdefW}
 \lambda^{(1)}(z)\:=\:\lambda^{(1)}+z\lambda^{(2)}, \qquad 
 \tilde{\lambda}^{(2)}(z)\:=\:\tilde{\lambda}^{(2)}-z\tilde{\lambda}^{(1)}.
\end{equation}
The on-shell representation induced by \eqref{BCFWdefW} shows just one pole at the location 
$z_{1n}\,=\,-\langle 1,n\rangle/\langle 2,n\rangle$ and it has the following form
\begin{equation}\eqlabel{MHVw}
 M_n\:=\:M_3(\hat{1}^{-},n^{+},-\hat{P}_{1n}^{+})\frac{f^{\mbox{\tiny $(\nu,n)$}}_{1n}}{P_{1n}^2}
         M_{n-1}(\hat{P}_{1n}^{-},(n-1)^{+},\ldots,4^{+},3^{-},\hat{2}^{+}).
\end{equation}
First of all, the large-$z$ behaviour is obtained through the condition in the last line of \eqref{zeroes-sum}. Being
characterised by just on term, the correct factorisation property as $\langle1,2\rangle\,\rightarrow\,0$ is obtained
for $\nu\,=\,\delta-2h_1\,=\,3$ and it is possible to see that there is just one zero of multiplicity $4$. Furthermore, the 
factor $\mathcal{H}_{n-1}$ is given by $\langle n,3\rangle^4/\langle P_{12},3\rangle^4$ and it has this form for any number
of external states. Thus, keeping in mind that from the identifications \eqref{PijId} $\lambda^{(12)}$ is nothing but 
$\lambda^{(2)}$, the last condition in \eqref{zeroes-sum} reduces to
\begin{equation}\eqlabel{MHVz}
 \lim_{\langle1,2\rangle\rightarrow0}\alpha_{1n}^{-4}\left(\frac{\langle n,3\rangle}{\langle 2,3\rangle}\right)^4\,=\,1,
\end{equation}
which is satisfied for
\begin{equation}\eqlabel{a1n}
 \alpha_{1n}\:=\:\frac{\langle n,3\rangle}{\langle 2,3\rangle}.
\end{equation}
The condition on the zeroes can be therefore written as
\begin{equation}\eqlabel{MHVzcond}
 P_{1n}^2(z_0)\:=\:\langle1,2\rangle\frac{\langle n,3\rangle}{\langle2,3\rangle}[1,n],\qquad\Rightarrow\qquad
 f_{1n}^{\mbox{\tiny $(3,n)$}}\:=\:
  \left(\frac{\langle1,3\rangle\langle2,3\rangle}{\langle1,2\rangle\langle n,3\rangle}\right)^{4}.
\end{equation}
Using the expression \eqref{MHVzcond} for $f_{1n}^{\mbox{\tiny $(3,n)$}}$, one can easily obtain the correct expression
for the tree-level MHV-amplitudes of gluons.

\subsection{Einstein-Maxwell Theory}\label{subsec:EM}

Let us consider now the scattering amplitudes of photon mediated by gravitons. In such a theory the relevant three-particle
amplitudes are of two types: photon-photon-graviton amplitude and the three-particle self-interaction for gravitons
\begin{equation}\eqlabel{EM3pt}
 M_{3}(a^{-1}_{\gamma},b^{+1}_{\gamma},c^{-2}_g)\:=\:\kappa\frac{\langle c,a\rangle^4}{\langle a,b\rangle^2},\qquad
 M_{3}(a^{-1}_{\gamma},b^{+1}_{\gamma},c^{+2}_g)\:=\:\kappa\frac{[b,c]^4}{[a,b]^2}.
\end{equation}
In this theory there is no BCFW-deformation involving two photons such that the amplitudes vanish at infinity in the 
large-$z$ limit. Amplitudes with external gravitons can satisfy standard BCFW-recursion relations, if the momenta of
two gravitons are deformed. We will analyse the cases in which the momenta of two photons are deformed:
\begin{equation}\eqlabel{EMdef}
 \tilde{\lambda}^{\mbox{\tiny $(1)$}}(z)\:=\:\tilde{\lambda}^{\mbox{\tiny $(1)$}}-z\tilde{\lambda}^{\mbox{\tiny $(2)$}},
  \qquad
 \lambda^{\mbox{\tiny $(2)$}}(z)\:=\:\lambda^{\mbox{\tiny $(2)$}}+z\lambda^{\mbox{\tiny $(1)$}}.
\end{equation}
Under the deformation \eqref{EMdef}, the analysis of the $[1,2]\,\rightarrow\,0$ leads to a condition of the form of
the third line in \eqref{zeroes-sum} which, for any $n$, implies $\nu\,=\,2h_1+\delta\,\equiv\,0$. Therefore, as 
$z\,\rightarrow\,\infty$, the amplitudes in this theory behave as a constant.

\subsubsection{Four-Particle Amplitudes}\label{subsubsec:EM4pt}

We start with constructing the amplitudes for this theory from the smallest non-trivial example: the four-particle 
amplitudes. There are two classes of four-particle amplitudes which need to be considered: 
$M_{4}(1^{-1}_{\gamma},2^{+1}_{\gamma},3^{-2}_{g},4^{+2}_g)$ and 
$M_{4}(1^{-1}_{\gamma},2^{+1}_{\gamma},3^{-1}_{\gamma},4^{+1}_{\gamma})$.
As far as the first amplitude is concerned, the on-shell representation \eqref{Mfin} shows just a single contribution
\begin{equation}\eqlabel{EMppgg}
 M_{4}(1^{-1}_{\gamma},2^{+1}_{\gamma},3^{-2}_{g},4^{+2}_g)\:=\:
  M_3(\hat{1}^{-1}_{\gamma},3^{-2}_g,-\hat{P}_{13}^{+1})\frac{f_{13}^{\mbox{\tiny $(0,4)$}}}{P_{13}^2}
  M_{3}(\hat{P}_{13}^{-1},4^{+}_g,\hat{2}^{+1}_{\gamma}),
\end{equation}
where:
\begin{equation}\eqlabel{EMhat}
 \hat{\tilde{\lambda}}^{\mbox{\tiny $(1)$}}\:=\:\frac{[2,1]}{[2,3]}\tilde{\lambda}^{\mbox{\tiny $(3)$}},\quad
 \hat{\lambda}^{\mbox{\tiny $(2)$}}\:=\:\frac{\langle1,2\rangle}{\langle1,4\rangle}\lambda^{\mbox{\tiny $(4)$}},\quad
 \hat{P}_{13}\:=\:\frac{\langle1,3\rangle}{\langle1,4\rangle}\lambda^{\mbox{\tiny $(4)$}}\tilde{\lambda}^{\mbox{\tiny $(3)$}}
 .
\end{equation}
For the moment, we pretend not to know that the zeroes satisfy the condition \eqref{4ptzeroes}, which will be discussed
in \cite{Benincasa:2011bc}. Analysing the limit $[1,2]\,\rightarrow\,0$ and considering that 
$\mathcal{H}_{3}^{\mbox{\tiny $(13)$}}\,=\,[4,3]^2/[P_{12},4]^2$ 
(where $\tilde{\lambda}^{\mbox{\tiny $(12)$}}\,=\,\tilde{\lambda}^{\mbox{\tiny $(2)$}}$), one obtains
\begin{equation}\eqlabel{EMz}
 \lim_{[1,2]\,\rightarrow\,0}\alpha_{13}^{-1}\frac{\langle1,3\rangle}{\langle1,2\rangle}\frac{[4,3]^2}{[2,4]^2}\,=\,1,
\end{equation}
which is satisfied for
\begin{equation}\eqlabel{EMa13}
 \alpha_{13}\:=\:\frac{\langle1,3\rangle}{\langle1,2\rangle}\frac{[4,3]^2}{[2,4]^2}
\end{equation}
Thus the condition on the zeroes reads
\begin{equation}\eqlabel{EMzcond}
 P_{13}^2(z_0)\:=\:-P_{12}^2,\qquad\Rightarrow\qquad f_{13}^{\mbox{\tiny $(0,4)$}}\:=\:-\frac{P_{14}^2}{P_{12}^2}.
\end{equation}
Using the expressions for the three-particle amplitudes \eqref{EM3pt} and of $f_{13}^{\mbox{\tiny $(0,4)$}}$ in
the on-shell representation \eqref{EMppgg}, the four-particle amplitude with two external photons and two external
gravitons turns out to be
\begin{equation}\eqlabel{EMppggFin}
  M_{4}(1^{-1}_{\gamma},2^{+1}_{\gamma},3^{-2}_{g},4^{+2}_g)\:=\:-\kappa^2
  \frac{\langle1,3\rangle^2\langle2,3\rangle^2 [4,2]^4}{s_{12}s_{13}s_{14}},
\end{equation}
where we used the Mandelstam variables 
$s_{ab}\,\overset{\mbox{\tiny def}}{=}\,(p^{\mbox{\tiny $(a)$}}+p^{\mbox{\tiny $(b)$}})^2$.

Let us now focus on the amplitudes with four external photons. As in the previous case, the on-shell representation
\eqref{Mfin} shows just one term
\begin{equation}\eqlabel{EMpppp}
 M_{4}(1^{-1}_{\gamma},2^{+1}_{\gamma},3^{-1}_{\gamma},4^{+1}_{\gamma})\:=\:
  M_{3}(\hat{1}^{-1}_{\gamma},4^{+1}_{\gamma},-\hat{P}_{14}^{-2})\frac{f_{14}^{\mbox{\tiny $(0,4)$}}}{P_{14}^2}
  M_{3}(\hat{P}_{14}^{+2},3^{-1}_{\gamma},\hat{2}^{+1}_{\gamma}).
\end{equation}
Proceeding in the same fashion, the analysis of the limit $[1,2]\,\rightarrow\,0$ leads to
\begin{equation}\eqlabel{EMa14}
 \alpha_{14}\:=\:\frac{\langle1,4\rangle}{\langle1,2\rangle}\frac{[4,3]^2}{[3,2]^2},
\end{equation}
so that the condition on the zeroes acquires the form
\begin{equation}\eqlabel{EMzcond2}
 P_{14}^2(z_0)\:=\:-P_{12}^2,\qquad\Rightarrow\qquad f_{14}^{\mbox{\tiny $(0,4)$}}\:=\:-\frac{P_{13}^2}{P_{12}^2}.
\end{equation}
Therefore, the four-photons scattering amplitude in Einstein-Maxwell theory is
\begin{equation}\eqlabel{EMppppFin}
 M_{4}(1^{-1}_{\gamma},2^{+1}_{\gamma},3^{-1}_{\gamma},4^{+1}_{\gamma})\:=\:
  -\kappa^2\frac{\langle1,3\rangle^2 [2,4]^2}{s_{12}s_{14}}s_{13}.
\end{equation}

\subsubsection{Five-Particle Amplitudes}\label{subsubsec:EM5pt}

We can continue to build the theory by computing higher-point amplitudes. In this sub-section we compute the $5$-particle 
amplitude $M_{5}(1^{-1}_{\gamma},2^{+1}_{\gamma},3^{-1}_{\gamma},4^{+1}_{\gamma},5^{-2}_g)$. For such an amplitude, we 
choose the on-shell representation obtained through the deformation \eqref{EMdef}, which shows just two terms
\begin{equation}\eqlabel{EM5pt}
 \begin{split}
  M_{5}(1^{-1}_{\gamma},2^{+1}_{\gamma},3^{-1}_{\gamma},4^{+1}_{\gamma},5^{-2}_g)\:=\:
  &M_{3}(\hat{1}^{-1}_{\gamma},4^{+1}_{\gamma},-\hat{P}_{14}^{-2})\frac{f_{14}^{\mbox{\tiny $(0,5)$}}}{P_{14}^2}
   M_{4}(\hat{P}_{14}^{+2},5^{-2}_g,3^{-1}_{\gamma},\hat{2}^{+1}_{\gamma})+\\
  &+M_{3}(\hat{1}^{-1}_{\gamma},5^{-2}_g,-\hat{P}_{14}^{+1})\frac{f_{15}^{\mbox{\tiny $(0,5)$}}}{P_{15}^2}
   M_{4}(\hat{P}_{15}^{-1},4^{+1}_{\gamma},3^{-1}_{\gamma},\hat{2}^{+1}_{\gamma}).
 \end{split}
\end{equation}
Requiring that $P_{1k}^2(z_0)\,=\,\langle1,k\rangle\alpha_{1k}[1,2]$, the condition induced by the limit 
$[1,2]\,\rightarrow\,0$ is
\begin{equation}\eqlabel{EM5pt12lim}
 \lim_{[1,2]\rightarrow0}
  \left[
    \alpha_{14}^{-1}\frac{\langle1,4\rangle}{\langle1,2\rangle}\mathcal{H}_{4}^{\mbox{\tiny $(4)$}}+
    \alpha_{15}^{-1}\frac{\langle1,5\rangle}{\langle1,2\rangle}\mathcal{H}_{4}^{\mbox{\tiny $(5)$}}
  \right]\:=\:-1.
\end{equation}
The factors $\mathcal{H}_4$'s can be easily computed from the knowledge of the four-particle amplitudes computed in
section \ref{subsubsec:EM4pt}, and they turn out to be
\begin{equation}\eqlabel{EM5ptHs}
 \mathcal{H}_{4}^{\mbox{\tiny $(k)$}}\,=\,\frac{[3,k]^2}{[3,P_{12}]^2},\qquad k\,=\,4,5.
\end{equation}
Considering that $\tilde{\lambda}^{(12)}\,=\,\tilde{\lambda}^{(2)}$, the condition \eqref{EM5pt12lim} is generically
satisfied for
\begin{equation}\eqlabel{EM5pta}
 \alpha_{14}\:=\:h_{14}\frac{\langle1,2\rangle}{\langle1,4\rangle}
  \left(\frac{\langle1,4\rangle[4,3]}{\langle1|P_{12}|3]}\right)^2,\qquad
 \alpha_{15}\:=\:h_{15}\frac{\langle1,2\rangle}{\langle1,5\rangle}
  \left(\frac{\langle1,5\rangle[5,3]}{\langle1|P_{12}|3]}\right)^2,
\end{equation}
with $h_{15}\,=\,1-h_{14}$. In order to completely fix the $\alpha_{1k}$'s, it is necessary to analyse other limits.
Specifically, the following conditions must hold:
\begin{equation}\eqlabel{EM5ptConds}
 \begin{split}
  &\hspace{2cm}\lim_{[3,4]\,\rightarrow\,0}f_{15}^{\mbox{\tiny $(0,5)$}}\:=\:f_{15}^{\mbox{\tiny $(0,4)$}},\qquad
   \lim_{[3,5]\,\rightarrow\,0}f_{14}^{\mbox{\tiny $(0,5)$}}\:=\:f_{14}^{\mbox{\tiny $(0,4)$}},\\
  &\lim_{[4,5]\,\rightarrow\,0}f_{14}^{\mbox{\tiny $(0,5)$}}\:=\:f_{1(45)}^{\mbox{\tiny $(0,4)$}}\:\equiv\:
    f_{23}^{\mbox{\tiny $(0,4)$}},\qquad
   \lim_{[4,5]\,\rightarrow\,0}f_{15}^{\mbox{\tiny $(0,5)$}}\:=\:f_{1(45)}^{\mbox{\tiny $(0,4)$}}\:\equiv\:
    f_{23}^{\mbox{\tiny $(0,4)$}},
 \end{split}
\end{equation}
where $f_{15}^{\mbox{\tiny $(0,4)$}}$, $f_{14}^{\mbox{\tiny $(0,4)$}}$ and $f_{23}^{\mbox{\tiny $(0,4)$}}$ are related to 
the four-particle amplitudes $M_4(1^{-1}_{\gamma},2^{+1}_{\gamma},P_{34}^{+2},5^{-2}_g)$, 
$M_4(1^{-1}_{\gamma},2^{+1}_{\gamma},P_{35}^{-1},4^{+1}_{\gamma})$ and 
$M_4(1^{-1}_{\gamma},2^{+1}_{\gamma},3^{-1}_{\gamma},P_{45}^{+1})$ respectively.
From the four-particle analysis in section \ref{subsubsec:EM4pt}, the channels computed at the location of the zeroes
are equal to $-P_{12}^2$:
\begin{equation}\eqlabel{EM5ptConds2}
 \begin{split}
  &\lim_{[3,4]\,\rightarrow\,0}P_{15}^2(z_0)\:=\:P_{15}^2(\bar{z}_0)\:=\:\langle1,5\rangle
    \frac{[5,P_{34}]}{[2,P_{34}]}[1,2]\:\equiv\:-P_{12}^2,\\
  &\lim_{[3,5]\,\rightarrow\,0}P_{14}^2(z_0)\:=\:P_{14}^2(\bar{z}_0)\:=\:\langle1,4\rangle
    \frac{[4,P_{35}]}{[2,P_{35}]}[1,2]\:\equiv\:-P_{12}^2,\\
  &\lim_{[4,5]\,\rightarrow\,0}\frac{P_{14}^2}{P_{14}^2(z_0)}\:=\:\frac{P_{23}^2}{P_{23}^2(z_0)}\:\equiv\:
    -\frac{P_{23}^2}{P_{12}^2},
 \end{split}
\end{equation}
where the $\bar{z}_0$ in the first line is the zero of the four-particle amplitude 
$M_4(1^{-1}_{\gamma},2^{+1}_{\gamma},P_{34}^{+2},5^{-2}_g)$, while the one in the second line is related to the 
four-particle amplitude $M_4(1^{-1}_{\gamma},2^{+1}_{\gamma},P_{35}^{-1},4^{+1}_{\gamma})$.
The conditions \eqref{EM5ptConds2} are satisfied if and only if 
\begin{equation}\eqlabel{EM5ptZeroes}
  P_{14}^2(z_0)\:=\:\langle1,4\rangle\frac{[4,3]}{[2,3]}[1,2],\qquad
  P_{15}^2(z_0)\:=\:\langle1,5\rangle\frac{[5,3]}{[2,3]}[1,2],\qquad
  \mbox{with: } h_{14}\:=\:\langle1,2\rangle\frac{[2,3]}{[4,3]}.
\end{equation}
Knowing the channels computed at the location of the zero, we have all the ingredients for computing the five-particle
amplitude from the on-shell representation \eqref{EM5pt}:
\begin{equation}\eqlabel{EM5ptFin}
 M_{5}\:=\:\kappa^3\frac{[1,3][2,4]^5
  \left(\langle1,4\rangle[4,3]\langle3,5\rangle[5,1]+\langle1,5\rangle[5,3]\langle3,4\rangle[4,1]\right)}{[1,2][1,4][1,5]
   [2,3][2,5][3,4][3,5][4,5]}.
\end{equation}

\section{Conclusion}\label{sec:Concl}

In this paper we extended the notion of tree-level constructibility to all theories of massless particles, through
the proof of the existence of a set of recursion relations
\begin{equation}\eqlabel{GenRR}
 M_{n}\:=\:\sum_{k\in\mathcal{P}^{\mbox{\tiny $(i,j)$}}}
  M_{\mbox{\tiny L}}^{\mbox{\tiny $(i,j)$}}(\hat{i},\mathcal{I}_k,-\hat{P}_{\mbox{\tiny $i\mathcal{I}_k$}})
  \frac{f_{\mbox{\tiny $i\mathcal{I}_k$}}^{\mbox{\tiny $(\nu,n)$}}}{P_{\mbox{\tiny $i\mathcal{I}_k$}}^2}
  M_{\mbox{\tiny R}}^{\mbox{\tiny $(i,j)$}}(\hat{P}_{\mbox{\tiny $i\mathcal{I}_k$}}, \mathcal{J}_{k}, \hat{j}),
\end{equation}
where the ``weight'' $f_{\mbox{\tiny $i\mathcal{I}_k$}}^{\mbox{\tiny $(\nu,n)$}}$ is given by
\begin{equation}\eqlabel{fweight2}
 f_{\mbox{\tiny $i\mathcal{I}_k$}}^{\mbox{\tiny $(\nu,n)$}}\:=\:
 \left\{
  \begin{array}{l}
   1,\hspace{4cm} \nu\,<\,0,\\
   \phantom{\ldots}\\
   \prod_{l=1}^{\nu+1}\left(1-\frac{P_{\mbox{\tiny $i\mathcal{I}_k$}}^2}{P_{\mbox{\tiny $i\mathcal{I}_k$}}^2
    \left(z_0^{\mbox{\tiny $(l)$}}\right)}\right),\quad \nu\,\ge\,0,
  \end{array}
 \right.
\end{equation}
generalising the known BCFW-recursion relations. This generalised on-shell representation for tree-level scattering 
amplitudes needs a new element: the knowledge of a subset of the zeroes of the amplitudes. To our knowledge, kinematic
limits where the scattering processes become trivial are not generically known. We argue that such kinematic points
can be fixed by imposing unitarity, {\it i.e.} that the representation \eqref{GenRR} shows the correct factorisation
properties in the collinear/multi-particle limits.

Specifically, we analyse in detail all the classes of collinear/multi-particle limits. Such an analysis allows to
show that the complex-UV (large-$z$) behaviour under a two-particle deformation depends only on the nature of the
interaction (through the number of its derivatives $\delta$) and on the helicities of the particles whose momenta are 
deformed. It is instead independent of the number of external states. This can be physically understood by interpreting this
complex-UV behaviour as a hard-particle limit, in which the deformed particles behave as a hard-particle of momentum
$zq$, while all the undeformed particles can be considered soft with respect to it \cite{ArkaniHamed:2008yf}. In such
a limit, the number of soft particles do not affect the leading behaviour in a large momentum $zq$ expansion.

In the on-shell representation \eqref{GenRR}, where $i$ and $j$ are the labels of the deformed momenta, the $(i,j)$-channel
is not contained explicitly. It was argued in \cite{Schuster:2008nh} that the correct factorisation in such a channel,
for amplitudes which admit a standard BCFW representation, is a consequence of the fact that the related singularity
appears as a soft singularity rather than as a collinear one. Furthermore, the standard BCFW construction fails when
such a singularity is actually not enough to reproduce the factorisation properties in this channel.

Requiring that \eqref{GenRR} factorises properly in this channel partially fixes the properties of the zeroes (or better,
of the channels computed at the location of the zeroes): the factors $f_{ik}^{\mbox{\tiny $(\nu,n)$}}$ and 
$f_{jk}^{\mbox{\tiny $(\nu,n)$}}$ allow for the correct singularity to show up in the $(i,j)$-channel. In principle, 
the channels computed at the location of the zeroes could be fixed by looking at the other collinear/multi-particle
limits. In particular, it seems to be possible to connect the factors $f_{i\mathcal{I}_k}^{\mbox{\tiny $(\nu,n)$}}$
to the ones of the lower point amplitudes. So far we were not able to find an exact relation - sort of recursive procedure -
to compute such factors and we leave it for future work. However, in many examples the knowledge of the conditions 
\eqref{zeroes-sum} is enough. Even if indeed this is not completely satisfactory, despite of this issue, 
the representation \eqref{GenRR} sheds light on the tree-level structure of theories of massless particles and on the
kinematic properties of the amplitudes. The latter indeed will be better understood once the general properties of the
factors $f_{i\mathcal{I}_k}^{\mbox{\tiny $(\nu,n)$}}$ are themselves better understood.

\section*{Acknowledgement}

It is a pleasure to thank Yacine Mehtar-Tani for valuable discussions. We also would like to thank Freddy Cachazo for 
reading the manuscript. This work is funded in part by MICINN under grant FPA2008-01838, by the Spanish Consolider-Ingenio 
2010 Programme CPAN (CSD2007-00042) and by Xunta de Galicia (Conseller{\'i}a de Educac{\'i}on, grant INCITE09 206 121 PR and
grant PGIDIT10PXIB206075PR) and by FEDER. P.B. is supported as well by the MInisterio de Ciencia e INNovaci{\'on} through 
the Juan de la Cierva program. E.C. is supported by a Spanish FPU fellowship, and thanks the FRont Of Galician-speaking 
Scientists for unconditional support.

\appendix

\bibliographystyle{utphys}
\bibliography{amplitudesrefs}

\providecommand{\href}[2]{#2}\begingroup\raggedright\begin{thebibliography}{10}

\bibitem{Benincasa:2007xk}
P.~Benincasa and F.~Cachazo, ``{Consistency Conditions on the S-Matrix of
  Massless Particles},''
\href{http://arxiv.org/abs/0705.4305}{{\tt arXiv:0705.4305 [hep-th]}}.

\bibitem{Britto:2004aa}
R.~Britto, F.~Cachazo, and B.~Feng, ``{New recursion relations for tree
  amplitudes of gluons},'' {\em Nucl. Phys. B} {\bf 715} (2005)  499,
  \href{http://arxiv.org/abs/0412308}{{\tt arXiv:0412308 [hep-th]}}.

\bibitem{Britto:2005aa}
R.~Britto, F.~Cachazo, B.~Feng, and E.~Witten, ``{Direct proof of tree-level
  recursion relation in Yang-Mills theory},'' {\em Phys. Rev. Lett.} {\bf 94}
  (2005)  , \href{http://arxiv.org/abs/0501052}{{\tt arXiv:0501052 [hep-th]}}.

\bibitem{Bedford:2005yy}
J.~Bedford, A.~Brandhuber, B.~J. Spence, and G.~Travaglini, ``{A recursion
  relation for gravity amplitudes},''
  \href{http://dx.doi.org/10.1016/j.nuclphysb.2005.016}{{\em Nucl. Phys.} {\bf
  B721} (2005)  98--110},
\href{http://arxiv.org/abs/hep-th/0502146}{{\tt arXiv:hep-th/0502146}}.

\bibitem{Benincasa:2007qj}
P.~Benincasa, C.~Boucher-Veronneau, and F.~Cachazo, ``{Taming tree amplitudes
  in general relativity},''
  \href{http://dx.doi.org/10.1088/1126-6708/2007/11/057}{{\em JHEP} {\bf 11}
  (2007)  057},
\href{http://arxiv.org/abs/hep-th/0702032}{{\tt arXiv:hep-th/0702032}}.

\bibitem{ArkaniHamed:2008yf}
N.~Arkani-Hamed and J.~Kaplan, ``{On Tree Amplitudes in Gauge Theory and
  Gravity},'' \href{http://dx.doi.org/10.1088/1126-6708/2008/04/076}{{\em JHEP}
  {\bf 04} (2008)  076},
\href{http://arxiv.org/abs/0801.2385}{{\tt arXiv:0801.2385 [hep-th]}}.

\bibitem{Cheung:2008dn}
C.~Cheung, ``{On-Shell Recursion Relations for Generic Theories},''
\href{http://arxiv.org/abs/0808.0504}{{\tt arXiv:0808.0504 [hep-th]}}.

\bibitem{ArkaniHamed:2008gz}
N.~Arkani-Hamed, F.~Cachazo, and J.~Kaplan, ``{What is the Simplest Quantum
  Field Theory?},''
\href{http://arxiv.org/abs/0808.1446}{{\tt arXiv:0808.1446 [hep-th]}}.

\bibitem{Witten:2003aa}
E.~Witten, ``{Perturbative gauge theory as a string theory in twistor space},''
  {\em Commun. Math. Phys.} {\bf 252} (2004)  189,
  \href{http://arxiv.org/abs/0312171}{{\tt arXiv:0312171 [hep-th]}}.

\bibitem{Feng:2009ei}
B.~Feng, J.~Wang, Y.~Wang, and Z.~Zhang, ``Bcfw recursion relation with nonzero
  boundary contribution,'' {\em JHEP} {\bf 01} (2010)  019,
  \href{http://arxiv.org/abs/0911.0301}{{\tt arXiv:0911.0301 [Unknown]}}.

\bibitem{Feng:2010ku}
B.~Feng and C.-Y. Liu, ``{A note on the boundary contribution with bad
  deformation in gauge theory},''
  \href{http://dx.doi.org/10.1007/JHEP07(2010)093}{{\em JHEP} {\bf 07} (2010)
  093},
\href{http://arxiv.org/abs/1004.1282}{{\tt arXiv:1004.1282 [hep-th]}}.

\bibitem{Bianchi:2008pu}
M.~Bianchi, H.~Elvang, and D.~Z. Freedman, ``{Generating Tree Amplitudes in N=4
  SYM and N = 8 SG},''
  \href{http://dx.doi.org/10.1088/1126-6708/2008/09/063}{{\em JHEP} {\bf 09}
  (2008)  063},
\href{http://arxiv.org/abs/0805.0757}{{\tt arXiv:0805.0757 [hep-th]}}.

\bibitem{Cheung:2009dc}
C.~Cheung and D.~O'Connell, ``{Amplitudes and Spinor-Helicity in Six
  Dimensions},''
\href{http://arxiv.org/abs/0902.0981}{{\tt arXiv:0902.0981 [hep-th]}}.

\bibitem{Boels:2009bv}
R.~Boels, ``Covariant representation theory of the poincar{\'e} algebra and
  some of its extensions,'' {\em JHEP} {\bf 01} (2010)  010,
  \href{http://arxiv.org/abs/0908.0738}{{\tt arXiv:0908.0738 [hep-th]}}.

\bibitem{CaronHuot:2010rj}
S.~Caron-Huot and D.~O'Connell, ``{Spinor Helicity and Dual Conformal Symmetry
  in Ten Dimensions},''
\href{http://arxiv.org/abs/1010.5487}{{\tt arXiv:1010.5487 [hep-th]}}.

\bibitem{Gang:2010gy}
D.~Gang, Y.-t. Huang, E.~Koh, S.~Lee, and A.~E. Lipstein, ``{Tree-level
  Recursion Relation and Dual Superconformal Symmetry of the ABJM Theory},''
  \href{http://dx.doi.org/10.1007/JHEP03(2011)116}{{\em JHEP} {\bf 03} (2011)
  116},
\href{http://arxiv.org/abs/1012.5032}{{\tt arXiv:1012.5032 [hep-th]}}.

\bibitem{Sugawara:1963aa}
M.~Sugawara and A.~Tubis, ``{Phase Representation of Analytic Functions},''
\href{http://dx.doi.org/10.1103/PhysRev.130.2127}{{\em Phys. Rev.} {\bf 130}
  (1963)  2127–2131}.

\bibitem{Jin:1964zz}
Y.~S. Jin and A.~Martin, ``{Connection Between the Asymptotic Behavior and the
  Sign of the Discontinuity in One-Dimensional Dispersion Relations},''
\href{http://dx.doi.org/10.1103/PhysRev.135.B1369}{{\em Phys. Rev.} {\bf 135}
  (1964)  B1369--B1374}.

\bibitem{Vernov:1975au}
Y.~S. Vernov, ``{The Properties of the Forward Elastic Scattering Amplitude at
  Asymptotic and Finite Energies},''
{\em Fiz. Elem. Chast. Atom. Yadra} {\bf 6} (1975)  601--631.

\bibitem{Zamiralov:1976qc}
V.~S. Zamiralov and A.~F. Kurbatov, ``{On Zeros of pi n Scattering
  Amplitude},''
{\em Yad. Fiz.} {\bf 23} (1976)  900--903.

\bibitem{Kurbatov:1977xb}
A.~F. Kurbatov, ``{Zeros of Scattering Amplitude},''
{\em Teor. Mat. Fiz.} {\bf 33} (1977)  354--363.

\bibitem{Vernov:1980xv}
Y.~S. Vernov and A.~F. Kurbatov, ``{Lower Bound and Zeros of the Scattering
  Amplitude},''
\href{http://dx.doi.org/10.1007/BF01029122}{{\em Theor. Math. Phys.} {\bf 43}
  (1980)  490--494}.

\bibitem{Schuster:2008nh}
P.~C. Schuster and N.~Toro, ``{Constructing the Tree-Level Yang-Mills S-Matrix
  Using Complex Factorization},''
\href{http://arxiv.org/abs/0811.3207}{{\tt arXiv:0811.3207 [hep-th]}}.

\bibitem{Benincasa:2011bc}
P.~Benincasa and E.~Conde, ``{Exploring the S-Matrix of Massless Particles},''
  {\em Work in progress}  .

\end{thebibliography}\endgroup

\end{document}